\documentclass[ preprint,pre,12pt]{revtex4-2}
\usepackage{amsfonts}
\usepackage{amssymb}
\usepackage{amsmath}
\usepackage{graphicx}
\usepackage{pgfplots}
\pgfplotsset{compat=1.18}

\begin{document}
\title{Semiclassical approximation  for barrier billiards}
\author{Eugene Bogomolny}
\affiliation{Universit\'e Paris-Saclay, CNRS,  LPTMS,  91405 Orsay, France}
\date{\today}
\begin{abstract}
Barrier billiards are simple examples of pseudo-integrable models which form an appealing  but poorly investigated subclass of dynamical systems. The paper examines the semiclassical limit of the exact quantum transfer operator for barrier billiards constructed in [J. Phys. A: Math. Theor. \textbf{55}, 024001 (2022)]. The obtained asymptotic expressions  are used to provide analytical arguments to support the conjecture that spectral statistical properties of barrier billiards are described by the semi-Poisson distribution   and to derive the trace formulas for such billiards which in the transfer operator approach is not automatic.  
 \end{abstract}

\maketitle

\section{Introduction}

The relationship between classical dynamics and statistical properties of corresponding quantum problems is the central topic of quantum chaos.  The principal conjectures in this field depict only  two limiting cases: (i) Quantum eigenenergies of typical classically integrable systems are statistically independent and  well described by the Poisson distribution  \cite{berry_tabor}, (ii) Quantum eigenenergies of generic classically chaotic systems are strongly correlated and their correlation functions agree with those of standard random matrix  ensembles depended only on system symmetries \cite{bgs, mehta}.These conjectures are well accepted and were numerically checked in a huge variety of different  problems.  

However there exist models which are neither integrable nor chaotic and, consequently,  they are not covered  by the above conjectures. A characteristic   example of such models is the so-called  pseudo-integrable billiards (see, e.g., \cite{richens_berry})  which are two-dimensional polygonal billiards whose all angles $\theta_j$ are rational multiplies of $\pi$. 

From numerous numerical calculations (see, e.g., \cite{karol}-\cite{xian}, besides others) it was observed that the spectral statistics of pseudo-integrable billiards differs from the above mentioned universal statistics but is similar to spectral statistical properties of the Anderson model at the metal-insulator point \cite{boris, shklovskii}. That type of statistics coined the name of  intermediate statistics and is characterised   by the following properties:
\begin{itemize}
\item Level repulsion as for standard  random matrix ensembles.
\item Exponential decrease of nearest-neighbour level distributions as for the Poisson statistics.
\item  Non-trivial value of the spectral compressibility.
\item Fractal properties of eigenfunctions (in the Fourier space).
\end{itemize}
Analytical results in this field are rare and  the majority of them is related with the calculation of the spectral compressibility by the summation over periodic orbits (cf., \cite{schmit}).  

\begin{figure}
\begin{center}
\includegraphics[width=.7\linewidth]{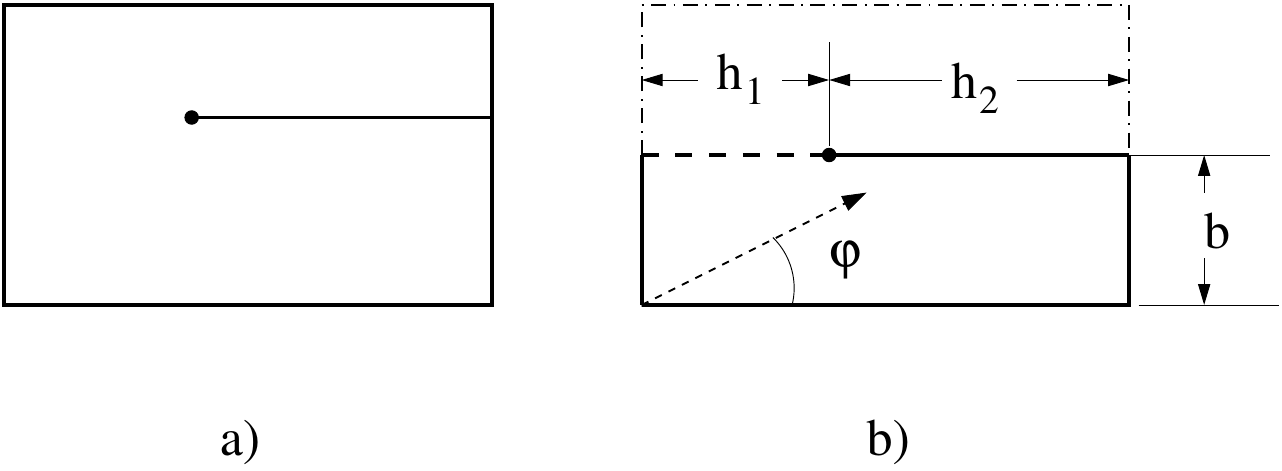}
\end{center}
\caption{(a) General barrier billiard. On all thick lines the Dirichlet boundary conditions are imposed.  
(b) Reduced symmetric barrier billiard. Dashed line is a part of the boundary with the Neumann boundary condition. Small black circle indicates, for clarity, the tip of the barrier. Dashed line with an arrow indicates schematically an initial segment of a  periodic orbit.}
\label{semiclassical_BB}
\end{figure}

Recently, a new analytical approach has been proposed in \cite{BB_I}-\cite{BB_III} to examine a specific example of pseudo-integrable billiards called barrier billiards which are rectangular billiards with a barrier inside (see Fig.~\ref{semiclassical_BB}(a)). The quantum problem consists in finding solutions of the Helmholtz equation
\begin{equation}
\left (\frac{\partial^2}{\partial x^2}+\frac{\partial^2}{\partial y^2}+k^2\right )\Psi(x,y)=0
\end{equation}
which obey the Dirichlet boundary conditions, $\Psi(x,y)=0$,  on the boundary of the rectangle and also along the barrier.  

The method used in \cite{BB_I, BB_II} to investigate barrier billiards consists of two parts. First, two  vertical billiard boundaries were removed and the problem was reduced to the scattering inside an infinite slab with a barrier parallel to slab boundary whose $S$-matrix  are  exactly obtained by the Wiener-Hopf method. Second, to force eigenfunctions to be zero at the vertical boundaries  special linear combinations of scattering solutions are constructed which, in the end, lead to the quantisation condition of the form
\begin{equation}
\det(1+B)=0
\label{quantisation}
\end{equation}    
where $B\equiv B(k)$ is an exact transfer operator (cf. \cite{transfer_op}) constructed from the scattering  $S$-matrix.

For simplicity,  only  symmetric barrier billiards  with a barrier  situated exactly at the middle of the rectangular are considered here. Due to the symmetry  non-trivial solutions of such billiard  correspond  to a rectangular billiard  with the height equal a half of the initial height but with different boundary conditions (partly Neumann, $\partial \Psi/\partial n=0$,  and partly Dirichlet, $\Psi=0$) along the upper boundary (see Fig.~\ref{semiclassical_BB}(b)).   Formulas for general barrier billiards are more cumbersome (cf. \cite{BB_II}) and will not be discussed here. 

For symmetric  barrier billiards the scattering  $S$-matrix has the form \cite{BB_I}: 
\begin{equation}
S_{n m}=\frac{L_n L_m}{x_n+x_m}
\label{S_matrix}
\end{equation}
where 
\begin{eqnarray}
& & x_{2m-1}=bp_{2m-1}\, ,\qquad x_{2m}=-bp_{2m}\, ,\qquad p_n=\sqrt{k^2-\frac{\pi^2}{4b^2}n^2}\, ,
\label{x_m}\\
& & L_{2n-1}=\frac{(-1)^n}{ \sqrt{b p_{2n-1}} K_{+}(p_{2n-1})}\, ,\qquad 
L_{2n}=\frac{(-1)^n \pi n \, K_{+}(p_{2n})}{\sqrt{b p_{2n}}}\, .  
\label{explicit_L}
\end{eqnarray}
As it is usual in the Wiener-Hopf method \cite{noble}, function  $K_{+}(\alpha)$ is the result of the factorisation (with $\mathrm{Im}\,k>0$)
\begin{equation}
\frac{\tan(q b)}{q b}=K_{+}(\alpha) K_{-}(\alpha),\qquad q=\sqrt{k^2-\alpha^2}
\end{equation}
such that  functions  $K_{\pm}(\alpha)$ are  free of   zeros and  singularities in, respectively,  the upper half-plane $\mathrm{Im}\alpha>-\mathrm{Im}\, k$  and in the lower half-plane $\mathrm{Im}\alpha< \mathrm{Im}\, k$.  A convenient explicit form of this function is \cite{BB_I} 
\begin{equation}
K_+(\alpha)=\prod_{n=1}^{\infty}\Big ( 1-\frac{1}{2n} \Big ) \Big (\frac{p_{2n}+\alpha}{p_{2n-1}+\alpha}\Big )\, .
\label{K}
\end{equation}
The transfer matrix $B$ in this case is 
\begin{equation}
B_{n m}(k)=e^{i\phi_n}S_{n m}\, ,\qquad \phi_{2n}=2h_2 p_{2n}\, ,\qquad \phi_{2n-1}=2h_1 p_{2n-1}\, . 
\label{B_matrix}
\end{equation}
Here $h_1$ and $h_2$ are, respectively, the lengths of the Neumann and the Dirichlet  parts of the boundary (see Fig.~\eqref{semiclassical_BB}(b)).  $h_1+h_2=a$ where $a$ is the length of the rectangle.

Though $K_{+}(\alpha)$ is given by an infinite (convergent) product, the moduli of $L_m$ are determined by  a finite product 
\begin{equation}
|L_m|^2=2x_m \prod_{j\neq m} \frac{x_m+x_j}{x_m-x_j}
\label{modulus_L}
\end{equation}
where $x_m$ are momenta of only propagating modes (i.e., with real momenta $p_m$)
and the matrix $B$ restricted to these propagating  modes is a finite unitary matrix of dimension $N=kb/\pi$  as it should be for transfer operators \cite{transfer_op}.   

The positivity of the right-hand side of expression \eqref{modulus_L} implies that real variables $x_m$ for propagating modes have to obey the following intertwining restrictions 
\begin{equation}
 |x_1|>|x_2|>\ldots >|x_N |\, ,\qquad x_m=(-1)^{m+1}|x_m| \, . 
 \label{inequalities}
 \end{equation}
 For $x_m$ as in \eqref{B_matrix}  these relations are automatic.  

It was argued in  \cite{BB_I} that phases $\phi_m$ in the $B$-matrix generically can be considered as independent random variables which makes this matrix a random unitary matrix The main result of \cite{BB_I}-\cite{BB_III} is that the spectral statistics of barrier billiards in the semiclassical limit $k\to\infty$ is independent of the position and the length of the barrier and is well described by the so-called semi-Poisson statistics with explicitly known correlation functions \cite{gerland} 
\begin{equation}
 P_n^{(\mathrm{sp})}(s)=\frac{2(2s)^{2n+1} e^{-2s}}{(2n+1)!}\, , \quad R_2^{(\mathrm{sp})}(s)=1-e^{-4s}\, ,
\quad  K^{(\mathrm{sp})}(\tau)=\frac{2+\pi^2 \tau^2}{4+\pi^2 \tau^2}\, , \quad \chi^{(\mathrm{sp})}=\frac{1}{2}\, . 
\label{semi_poisson}
\end{equation}  
Here $P_n(s)$ is the probability that two levels are separated by distance $s$ and inside this interval there exist exactly $n$ additional levels. $R_2(s)=\sum_{n=0}^{\infty} P_n(s) $ is the two-point correlation function
which determines the probability that two levels are at distance $s$ (with any numbers of levels inside). $K(\tau)$ is the two-point form factor defined as the Fourier transform of $R_2(s)$
\begin{equation}
K(\tau)=\int_{-\infty}^{\infty}(\delta(s)+R_2(|s|)-1)e^{2\pi i \tau s} ds
\end{equation}
and $\chi=K(0)$ is the spectral compressibility which determines the asymptotic behaviour of the variance of the number of levels, $n(L)$, in an interval $L$
\begin{equation}
\langle  (n(L)-\bar{n}(L))^2\rangle\underset{L\to\infty}{\longrightarrow} \chi L\,. 
\end{equation} 

The goal of this paper is to examine  the  barrier billiard transfer operator in the semiclassical limit  of high energies  and to investigate a few related problems.  Section~\ref{BB_semi_Poisson}  is devoted to the calculation of  the transfer operator when $k\to\infty$ and to the development of  analytical arguments  in favour of the statement that the spectral statistics of barrier billiards in this limit agrees with the semi-Poisson expressions.  In Section~\ref{trace_formula} the semiclassical trace formula for barrier billiards is derived  within the transfer operator  approach.  Surprisingly, the derivation is  not straightforward. Certain technical questions are discussed in Appendices.  
 Appendix~\ref{asymptotic} is devoted to the calculation of the infinite product  $K_+(\alpha)$ \eqref{K} in the limit $k\to\infty$ which permits to find the asymptotic expressions for all  matrix elements of the $S$-matrix \eqref{S_matrix}. In Appendix~\ref{images} it is shown that these asymptotic values can be obtained by  the direct summation of all images of the Sommerfeld  diffraction on a half-plane.    Properties of barrier billiard periodic orbits  are studied  in Appendix~\ref{ppo}. 

\section{Barrier billiard and semi-Poisson statistics}\label{BB_semi_Poisson}

Numerical calculations performed in \cite{BB_I,BB_II} for the unitary $B$-matrix \eqref{B_matrix} with   random phases $\phi_m$  and different choices of  real coordinates $x_m$ obeying \eqref{inequalities} (and, in particular, the ones specific for  barrier billiards \eqref{x_m})   have demonstrated that spectral statistics of such  $B$-matrices are well described by the semi-Poisson distribution but it has not  yet  been confirmed  analytically  (except the calculation of the spectral compressibility in \cite{BB_III}). 

In this Section three different random unitary matrices are considered. These matrices were chosen in such a way that they have the same form in the limit of large  matrix dimension $N\to\infty$.   One of these matrices is the $B$-matrix \eqref{B_matrix} for propagating modes which describes the spectral statistics of barrier billiards.

 The second unitary random matrix called below the $A$-matrix  is 
\begin{equation}
A_{nm}=e^{ i \phi_n}  \Sigma_{nm}\, ,\qquad \Sigma_{nm}= \frac{1}{N \cos(\pi (n-m)/N)}\, , \qquad n,m=1,\ldots, N\, ,\qquad N=\mathrm{odd}  
\label{A_matrix}
\end{equation} 
is a rare example where one can prove that its statistics at large $N$ is the semi-Poisson one as it is a reduction of a random Lax matrix for an integrable Ruijsenaars-Schneider model \cite{lax_matrices, integrable_ensembles}. 

The third matrix called the $C$-matrix  is a block matrix ($N=2N_0$)  of the following form 
\begin{equation}
 C_{nm}=e^{i \phi_n}Z_{nm}, \qquad Z_{nm}= \left ( \begin{array}{cc}  0& z\\ z^{ \mathrm{T}} & 0 \end{array}\right ), \qquad  z_{jk}=\frac{(-1)^{j+k} }{N_0\sin(\pi  (j-k-1/2)/N_0)}  .
\label{block_C}
\end{equation}
 Matrix $e^{i\phi_j}z_{jk}$  is also a certain reduction of the Lax matrix of a Ruijsenaars-Schneider model \cite{lax_matrices, integrable_ensembles}  but  its spectral statistics is very different from statistics of matrix $\Sigma_{nm}$. 
 
These three ensembles of random matrices have certain common points. First, they all are finite unitary matrices obtained by the multiplication of constant matrices by random phases. Second,  without random phases all these matrices are symmetric real unitary matrices  which implies that their eigenvalues are $\pm 1$. Third, and the most important for our purposes, is that  their limits $N\to\infty$ exist and for all of them have the same block form
 \begin{equation}
 A,B,C\underset{N\to\infty}{\longrightarrow} e^{i\phi_m} \left ( \begin{array}{cc}  0& s\\ s^{ \mathrm{T}} & 0 \end{array}\right )\, , \qquad  s_{jk}= \frac{(-1)^{j+k}}{\pi(j-k+1/2) }\, .
 \end{equation}  
Such asymptotic equivalence together with certain additional considerations makes it reasonable that local spectral statistics of all three matrices in the limit $N\to\infty$ are the same. As one of them is known to be described by the semi-Poisson distributions it follows (at least heuristically) that the two others (and, in particular,  the barrier billiards) also have the same statistics.   

\subsection{Paraxial approximation}\label{paraxial_approximation}

To find the transfer operator  in the limit $k\to\infty$ it is necessary to know the asymptotic value of $K_{+}(\alpha)$. This  function is given by the  infinite convergent product \eqref{K} and standard calculations presented in Appendix~\ref{asymptotic} proved that when $\alpha=\mathcal{O}(k)$ 
\begin{equation}
K_{+}(\alpha)\underset{k\to\infty}{\longrightarrow} \frac{e^{i\pi/4}}{\sqrt{b(k+\alpha)}}\, .
\label{K_asymptotic_II}
\end{equation}
From  \eqref{explicit_L} it is possible to  find all matrix elements of the scattering $S$-matrix \eqref{S_matrix} (see \eqref{s_2n-1_2m}-\eqref{s_2n_2m}). 

The key  limit corresponds to the case (called here the paraxial approximation)  when longitudinal momenta $p_{2m}, \,p_{2n-1}$  are large but their difference is small.  It means that if $p_{2m}=k\cos(\phi_1)$ and $p_{2n-1}=k\cos(\phi_2)$ with $\phi_j$ not too close to $\pi/2$  then $k (\phi_1-\phi_2)=\mathcal{O}(1)$.  

From the structure of the $S$-matrix \eqref{S_matrix} and the definition of $x_m$ it follows that 
under such conditions $S_{2m,2n-1}$ and $S_{2m-1,2n}$ are $\mathcal{O}(1)$ but  $S_{2m,2n}$ and $S_{2m-1,2n-1}$ are $\mathcal{O}(1/k)$.  Physically it means that we ignore  the reflected waves and take into account  only the transmitted waves. 

From \eqref{s_2n-1_2m}  one reads
\begin{equation}
S_{2n-1,2m}=\frac{(-1)^{n+m}\, \pi m\,  \sqrt{ k+p_{2n-1}}}{b^2 \sqrt{p_{2n-1} p_{2m}}(p_{2n-1}-p_{2m})\, \sqrt{k+p_{2m}}}\, . 
\end{equation}
The paraxial approximation consists in the assumption  that $p_{2m}\approx p_{2m-1}$.  Then
\begin{equation}
p_{2n-1}-p_{2m}=\frac{\pi^2(m^2-(n-1/2)^2)}{b^2(p_{2n-1}+p_{2m})}\approx \frac{\pi^2( m^2-(n-1/2)^2)}{2b^2 p_{2m}}\, . 
\end{equation}
Therefore in such approximation 
\begin{equation}
S_{2n-1,2m}^{(\mathrm{pa})}=\frac{(-1)^{n+m}}{\pi} \left ( \frac{1}{m-n+1/2}+\frac{1}{m+n-1/2} \right )\, .
\label{asymptotic_S}
\end{equation}
As this result corresponds to small-angle scattering, it may  also be obtained  by an analog of the Fraunhofer diffraction.  It means that this $S$-matrix can be calculated  simply by re-expansion of old (initial) wave functions into series of new functions (see also \cite{BB_III}). For symmetric  barrier billiards of width $b$ initial (resp. final) normalised elementary solutions propagating in a slab with Neumann - Dirichlet (resp. Dirichlet - Dirichlet) boundary conditions are 
\begin{equation}
\phi^{(+)}_{2n-1}(x,y) =\frac{e^{i p_{2n-1}x}}{\sqrt{b p_{2n-1}}} \sin\Big( \frac{\pi (n-1/2)}{b}y \Big)\, ,\qquad 
\phi^{(+)}_{2n}(x,y) =\frac{e^{i p_{2n}x}}{\sqrt{b p_{2n}}} \sin\Big( \frac{\pi n}{b}y \Big)\, . 
\label{elementary_solutions}
\end{equation} 
Therefore in the paraxial approximation the $S$-matrix are determines by the following expansion
\begin{equation}
\sin\Big( \frac{\pi (2n-1)}{2b}y \Big)=\sum_{m=1}^{\infty} S_{2n-1,2m}^{(\mathrm{pa})} \sin\Big( \frac{\pi 2m}{2b} y\Big)\, . 
\end{equation}
Here it is  taken  into account that in such (paraxial)  approximation longitudinal momenta are considered equal $p_{2m}\approx p_{2n-1}$. In particular, it means that the normalisation of  all solutions are the same. 
 
The necessary integral is elementary   
\begin{equation}
S_{2n-1,2m}^{(\mathrm{pa})}=\frac{2}{b} \int_{0}^b  \sin\Big( \frac{\pi (2n-1)}{2b}\Big )\sin\Big( \frac{\pi 2m }{2b} y\Big)dy=\frac{2(-1)^{m+n} m}{\pi \big (m^2-(n-1/2)^2\big)}
\end{equation}
which fully agrees with \eqref{asymptotic_S}. 

One can further simplify this matrix by taking into account only the pole term 
 \begin{equation}
 S_{2n-1,2m}^{(\mathrm{pa})} \approx \frac{(-1)^{n+m}}{\pi(m-n+1/2) }\, .
 \label{S_paraxial}
 \end{equation}
 Let us re-arrange matrix elements of the $S$-matrix by  grouping together indices of the same parity  so the full matrix has four blocks structure
  \begin{equation}
S=\left ( \begin{array}{cc}  S_{2n-1,2m-1}& S_{2n-1,2m}\\ S_{2m,2n-1} & S_{2m,2n}  \end{array}\right )\, .
\end{equation}
As the $S$-matrix is symmetric (the reciprocity principle) it means that in the paraxial approximation only elements $S_{2m-1,2n}=S_{2n, 2m-1}$ are non-zero in the limit $N\to\infty$  (or $k\to\infty$) and the full $S$-matrix has the form  
 \begin{equation}
S^{(\mathrm{pa})}=\left ( \begin{array}{cc}  0& s\\ s^{ \mathrm{T}} & 0 \end{array}\right )\, , \qquad  s_{jk}= \frac{(-1)^{j+k}}{\pi(k-j+1/2) } \, .
\label{block_S}
\end{equation}
This limiting paraxial $S$-matrix is formally unitary as it comes from an expansion of one set of orthogonal functions into another set of orthogonal functions but  the unitarity is achieved only  provided the summation over intermediate states is done over all integers 
\begin{equation}
\sum_{k=-\infty}^{\infty} s_{n k}s_{m k}=\delta_{nm}\,.
\end{equation}
It can be confirmed by usual formulas (with the symmetric summation over $k$)
\begin{equation}
\sum_{k=-\infty}^{\infty} \frac{1}{x-k}=\pi \cot(\pi x)\, ,\qquad \sum_{k=-\infty}^{\infty} \frac{1}{(x-k)^2}=\frac{\pi^2}{\sin^2 \pi x}\, .
\end{equation}
When matrix indices are finite the matrix is not unitary and its eigenvalues are not on the  unit circle which makes the investigation of  spectral statistics of the $B$-matrix \eqref{B_matrix} in the paraxial approximation meaningless.  

It is well understood that the linear decrease  of matrix elements from the diagonal (as in \eqref{S_paraxial})  is the principal  ingredient of random matrix models with intermediate statistics \cite{boris, shklovskii}.  Therefore it is natural to conjecture (at least heuristically)  that different models  which have the same asymptotic behaviour of matrix elements should have the same local spectral statistics. Thus, to elucidate  spectral correlation functions of the 
$B$-matrix (i.e., barrier billiards)  one could try to find a different matrix model with known statistical properties which have the same asymptotic as in \eqref{block_S}.

\subsection{Model A} 

Let us consider a seems to be unrelated matrix 
\begin{equation}
L_{nm}=e^{i\phi_n} M_{nm}, \qquad n,m=1,\ldots,N
\label{matrix_L}
\end{equation}
where matrix $M_{nm}$ depends on one free real parameter $\alpha$
\begin{equation}
M_{nm}=W_n(\theta)  \frac{ \sin (\pi \alpha ) }{\sin \left ( \frac{\theta_n-\theta_m}{2}+\pi \alpha \right ) } V_m(\theta)\, .
\label{general_M_matrix}
\end{equation}
Here 
\begin{equation}
W_n^2(\theta) =\prod_{k\neq n} \frac{ \sin \left ( \frac{\theta_n-\theta_k}{2} +\pi \alpha\right )}{\sin \left ( \frac{\theta_n-\theta_k}{2}\right )}\, ,\qquad 
V_n^2(\theta) =\prod_{k\neq n} \frac{ \sin \left ( \frac{\theta_n-\theta_k}{2} -\pi \alpha\right )}{\sin \left ( \frac{\theta_n-\theta_k}{2}\right )}\, .
\end{equation}
It appears that this matrix is the Lax matrix for a certain integrable compact Ruijsenaars-Schneider model \cite{lax_matrices, integrable_ensembles} such that $\phi_j$ are momenta of $N$ particles and $\theta_j$ are their coordinates. This matrix is meaningful provided that coordinates $\theta_n$ are such that  $W_n^2(\theta)$ and $V_n^2(\theta)$ are of the same signs for all indices. Denote this allowed region of $\theta$  by  $\Omega_N (\theta)$. When $\theta\in \Omega_N(\theta)$ matrices \eqref{general_M_matrix} and \eqref{matrix_L} are unitary matrices. 

 Due to the existence of integrable structure of action-angle variables  in the considered Ruijsenaars-Schneider model \cite{ruijsenaars_I}-\cite{ruijsenaars_III}  it is possible to prove  \cite{lax_matrices, integrable_ensembles}  that if phases (momenta) are uniformly distributed in the interval $(0,2\pi)$ and $\theta_j$ are uniformly distributed in the region $\Omega_N(\theta)$ then eigenvalues of matrix $L$ in \eqref{matrix_L} (denoted below by $\lambda_{\alpha}$)  are also uniformly distributed in $\Omega_N(\lambda)$. It means that the joint probability of eigenvalues of matrix \eqref{matrix_L}   has the form 
\begin{equation}
dP(\lambda_1,\ldots,\lambda_N)=C \prod_{\lambda_j \in \Omega_N(\lambda) } d\lambda_j 
\label{eigenvalue_distribution}
\end{equation}
with a normalisation constant $C$.

Contrary to  the usual random matrix ensembles, no wronskians are presented in this expression.  Nevertheless, the eigenvalue  spectral statistics of such models  is non-trivial due to the structure of allowed region $\Omega_N(\lambda)$.  The explicit form of this variety  depends strongly on $\alpha$ and $N$ \cite{lax_matrices, integrable_ensembles}. The properties of only two special cases will be reminded here. For other cases consult  \cite{lax_matrices, integrable_ensembles}.   

The first case corresponds to 
\begin{equation}
\alpha=\frac{1}{2}\, .
\label{first_alpha}
\end{equation}
 For this value the  $M$-matrix  takes the following form
\begin{equation}
M_{nm}=\frac{l_n l_m}{\cos((\theta_n-\theta_m)/2)}\, ,\qquad |l_m|^2 =(-1)^{(N-1)/2}\prod_{j\neq m}\cot ((\theta_m-\theta_j)/2)\, .
\label{M_matrix}
\end{equation}
It is easy to check that such matrix can be obtained from the above $S$-matrix \eqref{S_matrix} after the  substitution  instead of real $x_m$ complex  $y_m=e^{i\theta_m}$ with real $\theta_m$ (an analog of the Wick rotation). 

This matrix is unitary provided  $\theta_j$ obey  intertwining conditions similar to \eqref{inequalities}.  
Let $z_j$ be an ordered set of $N$ numbers with odd $N$ 
\begin{equation} 
0=z_1< z_2<\ldots<z_N< \pi
\label{z_j}
\end{equation}
 then the allowed region of $\theta_j$ where matrix \eqref{M_matrix} is unitary is defined as follows 
\begin{equation}
\Omega_n(\theta): 
 \theta_j =\left \{ \begin{array}{cc} z_j, &j=\mathrm{odd}\\ z_j+\pi, & j=\mathrm{even}\end{array}
 \right . \, .
 \label{omega_N} 
\end{equation}
The simplicity of this region permits to find all local correlation functions of eigenvalue distribution \eqref{eigenvalue_distribution} analytically even at finite (but odd) $N$.  For example, the nearest-neighbour distribution, $P_0(x)$, is the probability that two eigenvalues are at distance $x$ and there is no eigenvalues inside this interval.   According to the definition of $\Omega_N$ (see \eqref{omega_N}) between any two near-by eigenvalues  there exists another eigenvalue whose symmetric  image has to  be in-between  these two. The integral of it is, evidently, $x$. All other eigenvalues (together with symmetric images of even levels) should be ordered and to be  in-between $x$ and $\pi$. Therefore the integration over them is proportional to $(\pi-x)^{N-3}$. Rescaling the eigenvalues to mean level density $s=(N/2\pi) x$ and impose the standard normalisations 
one proves that  the nearest-neighbour distribution for matrix $L$ in \eqref{matrix_L} with finite $N$ is 
\begin{equation}
P_0(N, s)=\frac{2^2}{N^2}C_{N-1}^{2} s\Big (1-\frac{2s }{N}\Big )^{N-3}\, . 
\label{P_0_N}
\end{equation}
Similar considerations  prove that the probability, $P_n(N,s)$, that in the interval $0<s<N/2$ there exist exactly $n\geq 0$ eigenvalues ($N\geq 2n+3$)
\begin{equation}
P_n(N, s)=\Big (\frac{2}{N}\Big )^{2n+2} 2(n+1) C_{N-1}^{2n+2} s^{2n+1}
 \Big (1-\frac{2s}{N} \Big )^{N-2n-3} 
 \label{P_n_A}
\end{equation}
where $C_{m}^{n}$ are binomial coefficients. 

\begin{figure}
\begin{minipage}{.49\textwidth}
\begin{center}
\includegraphics[width=.9\linewidth]{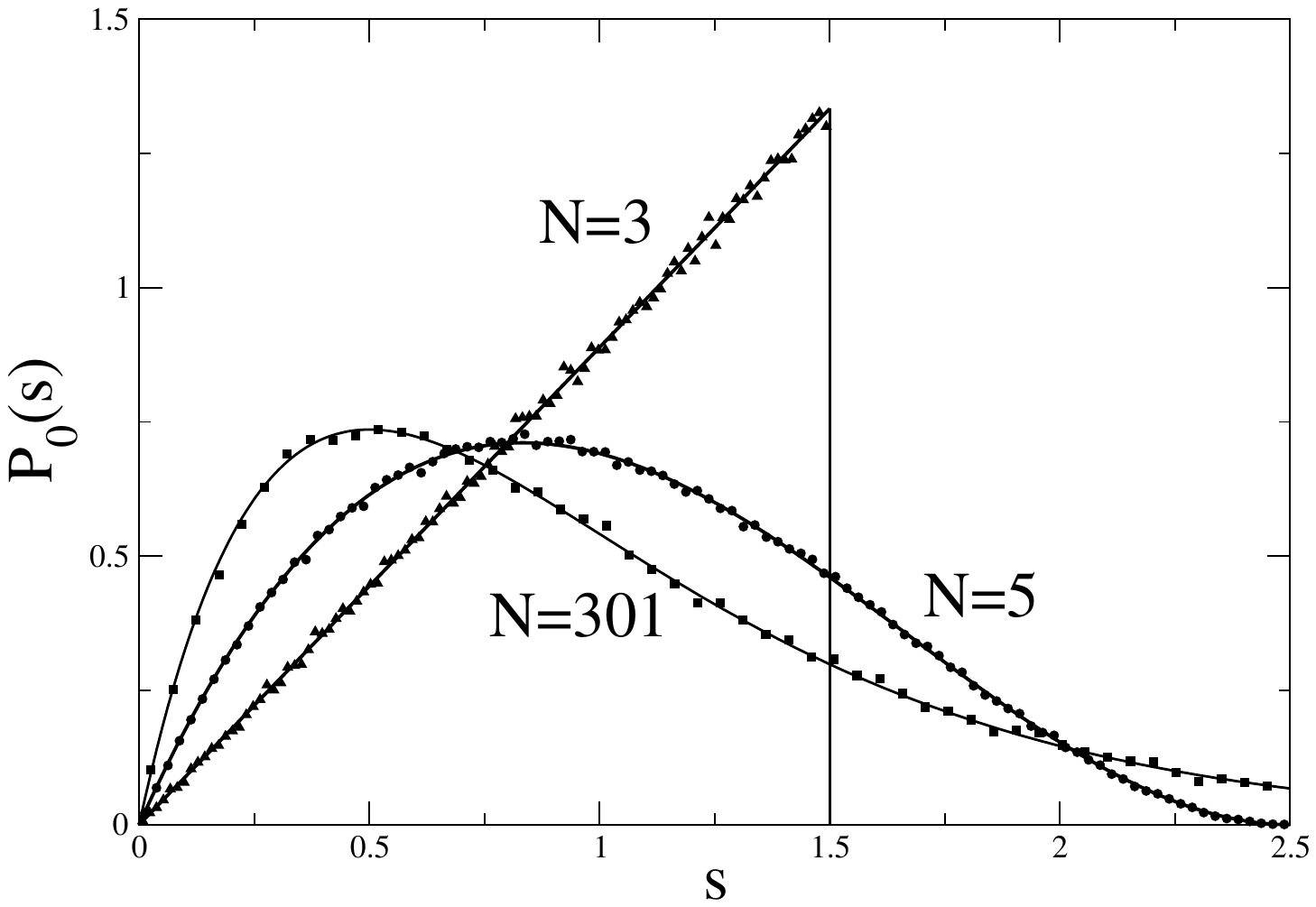}\\
a)
\end{center}
\end{minipage}
\begin{minipage}{.49\textwidth}
\begin{center}
\includegraphics[width=.9\linewidth]{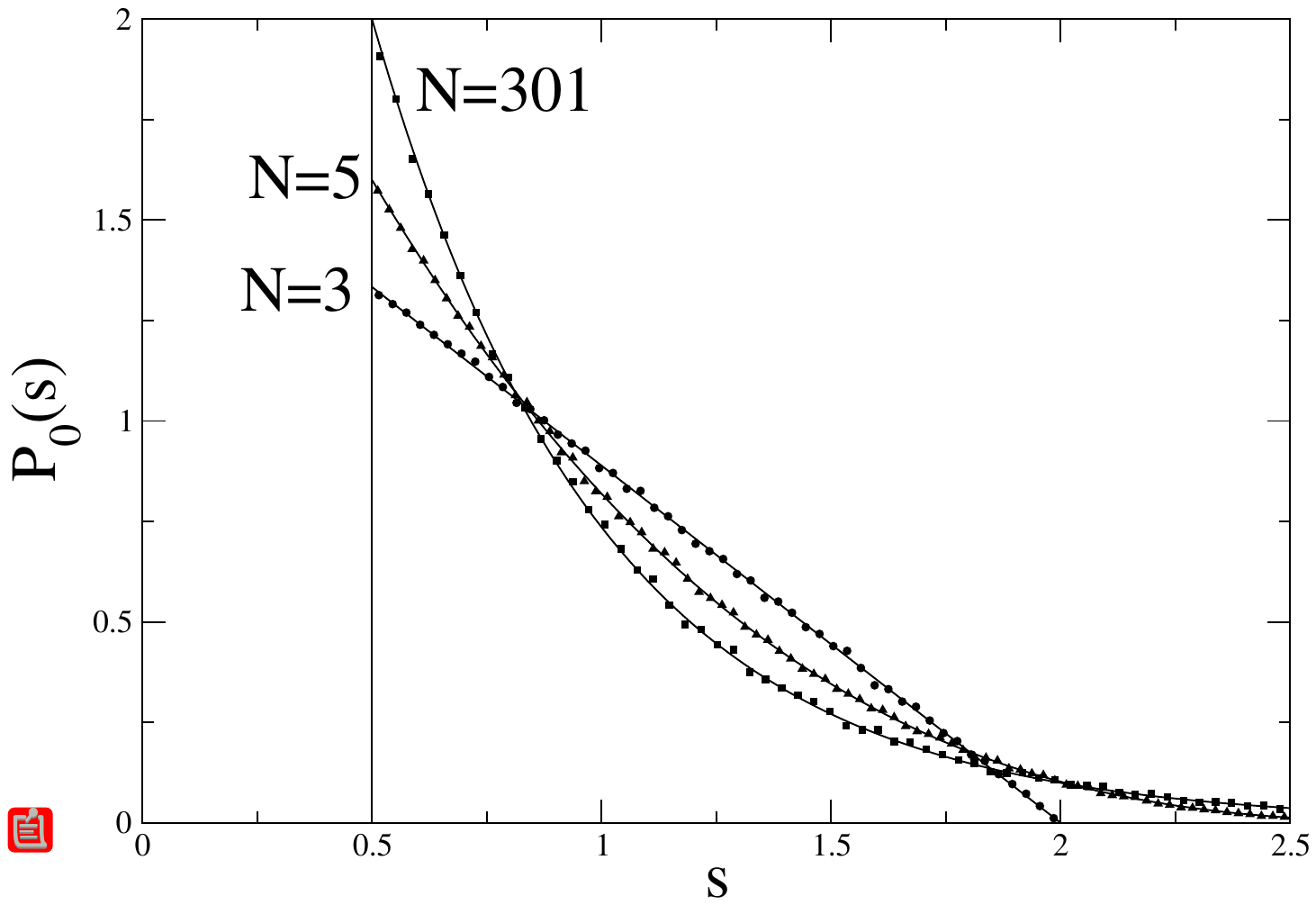}\\
b)
\end{center}
\end{minipage}
\caption{Nearest-neighbour distribution for integrable random  matrix ensembles \eqref{matrix_L} with uniform  distribution of  coordinates and momenta. (a) corresponds to $\alpha=\frac{1}{2}$ and (b) to $\alpha=\frac{1}{2N}$.  Triangles correspond to $N=3$, circles to  $N=5$, and squares to $N=301$. In the first two cases  $100\,000$ realisations were calculated, in the third $300$ realisations were taken into account. Solid lines are exact distributions  \eqref{P_n_A} and \eqref{P_n_C} for these values of $N$. In (a)  $P_0(3,s)=\frac{8}{9}s$, $P_0(5,s)=\frac{48}{25}s\left (1-\frac{2}{5}s\right )^2$ with $0<s<\frac{5}{2}$, and $P_0(301,s)\approx  P_0^{(\mathrm{sp})}(s)=4s\exp(-2s)$.  In (b) $P_0(3,s)=\frac{16}{9}\left (1-\frac{s}{2}\right )$, 
$P_0(5,s)=\frac{1728}{525}\left ( 1-\frac{s}{3} \right )^3$, $P_0(300,s)\approx P_0^{(\mathrm{shift\;P}) }(s)=2e^{-2s+1}$. }
\label{fig_old_map}
\end{figure}

When $N\to\infty$ with fixed $n$ these distributions  tends to the known  semi-Poisson expressions 
\eqref{semi_poisson}. For illustration, in Fig.~\ref{fig_old_map}(a) the results of numerical calculations for $N=3$, $5$, and $301$ are presented. It is clearly seen that the numerics  confirm  well the above formulas. The calculations were done in the following manner. First, $N-1$ points were independently and uniformly chosen in $(0,\pi)$. Then the points were numerated according to ascendent order (cf. \eqref{z_j}). Finally, every even points were shifted by $\pi$ as required by \eqref{omega_N}.  

The $B$-matrix \eqref{B_matrix} which describes the barrier billiard also corresponds to a  
Ruijsenaars-Schneider model but to a non-compact one. No 'natural' compactification of this model  such that the resulting  correlation functions are calculated analytically has not  been found yet. 

 There exist several variants of the random integrable model  \eqref{M_matrix} \cite{integrable_ensembles}.   Using the condensation of measure theory (see, e.g., \cite{condensation_measure}) one can argue that for large $N$ almost all (but fixed) realisations of coordinates $\theta_m$ produce the $L$-matrix \eqref{matrix_L} whose spectral statistics is well approximated by the semi-Poisson distributions \eqref{semi_poisson}. The simplest  choice is 
 \begin{equation}
 \theta_j=\frac{2\pi j}{N} \, ,\qquad j=0,1,\ldots,N-1
 \end{equation}
for which conditions \eqref{z_j} and \eqref{omega_N} are automatically satisfied. Simple calculations show that for these coordinates $l_m^2=1/N$ and matrix $M_{nm}$ denote, for clarity, by $\Sigma_{nm}$ takes a particular simple form
\begin{equation}
\Sigma_{nm}= \frac{1}{N \cos(\pi (n-m)/N)}\, .
\end{equation}
After the multiplication by random phases one gets model A  indicated in \eqref{A_matrix}. 

This matrix has appeared as  a particular example of a quantisation of  an interval-exchange map \cite{map} and has been investigated in \cite{interval_exchange}.   Such matrix with i.i.d. random phases, $\phi_m\in [0,2\pi)$, is the simplest example of  random matrix ensemble with the semi-Poisson statistics in the limit of large matrix dimension. Similar to the $S$-matrix, matrix $\Sigma$ is a finite real unitary (i.e., orthogonal) symmetric matrix such that $\Sigma^2=1$.    

Consider now  matrix  \eqref{A_matrix} in the limit $N\to\infty$. It is clear that non-zero matrix elements of the $\Sigma$-matrix are close to two diagonals $n-m=\pm [N/2]$. As $N$ has to be odd it means that  formally   
 \begin{equation}
\Sigma\underset{N\to\infty}{\longrightarrow} \left ( \begin{array}{cc}  0& s\\ s^{ \mathrm{T}} & 0 \end{array}\right )\, , \qquad  s_{jk}= \frac{(-1)^{j+k}}{\pi(k-j+1/2) }\, .
\label{block_U}
\end{equation}
As it is the same as the limit $N\to\infty$  of the $S$-matrix (cf. \eqref{block_S}) one may conclude that spectral statistics of the $B$-matrix  \eqref{B_matrix} in the limit of large $N$ is well described by spectral statistics of the $A$-matrix \eqref{A_matrix}, i.e., the semi-Poisson statistics \eqref{semi_poisson}.  This conclusion is based on the above (heuristic) conjecture that matrices with the the same asymptotic  linear falloff of matrix elements have the same spectral properties in the limit of large matrix dimensions.  

Strictly speaking, the above reasoning could be applied only for odd $N$. The case of  even matrix dimension requires modifications.  It is plain that many unitary matrices may have the same asymptotic behaviour of matrix elements as  in \eqref{block_S}.

\subsection{Model C}

Consider again the matrix  \eqref{general_M_matrix} but with a different value of parameter $\alpha$ 
\begin{equation}
\alpha=\frac{1}{2N}\, . 
\label{second_alpha}
\end{equation} 
It was shown in \cite{integrable_ensembles} that in this case the allowed region $\Omega_N(\theta)$ consists  of $N$ points on a circle such that the differences between near-by points are  larger than $\pi/N$. A convenient representation of this region is the following
\begin{equation}
\Omega_N(\theta)\;: \;0=\theta_1<\theta_2<\ldots \theta_N<2\pi -\pi/N ,\qquad \theta_{j+1}-\theta_j>\frac{\pi}{N}\, . 
\end{equation}
The joint probability of these points is given by \eqref{eigenvalue_distribution} with the above definition of the allowed region. It is plain  that correlation functions in such case correspond to the Poisson distribution but for 'thick' points of the size $\pi/N$. 

Consider $N$ independent points uniformly distributed at interval of length $L$. The probability that in  an interval $x$ there are exactly $n$ points is given by the known Poisson distribution
\begin{equation}
P_n^{(\mathrm{p})}(x)=\frac{N-1}{L^{N-1}}C_{N-2}^n   (L-x)^{N-2-n}x^n\, , \qquad 0<x<L \, .
\end{equation}
For the model $C$ eigenvalues of matrix   \eqref{general_M_matrix}  are uniformly distributed on the unit circle and after the unfolding the distribution of its eigenvalues are obtained from the above Poisson formulas with the shift: $L\to N(1-1/2)$ and $x\to s-(n+1)/2$. In the end one gets that  the nearest-neighbour distributions for this model is  
\begin{equation}
P_n^{(\mathrm{C})}(N, s)=\frac{2^{N-1}(N-1)}{N^{N-1}}C_{N-2}^n\left (\frac{N+n+1}{2}-s\right )^{N-2-n}\left ( s-\frac{n+1}{2}\right )^n 
\label{P_n_C}
\end{equation}
with the following restriction 
\begin{equation}
\frac{n+1}{2} < s < \frac{N+n+1}{2}\, .
\end{equation}  
When $N\to\infty$ these nearest-neighbour distributions tend to the shifted Poisson distribution
\begin{equation}
P_n^{(\mathrm{shift\;P})}(s)=\left \{ \begin{array}{cc} 0, & 0<s<(n+1)/2 \\  
\frac{2 (2s-n-1))^n }{n!} e^{-(2s-n-1))}, & s>(n+1)/2 \end{array}\right .\,  .
\label{shifted_poisson}
\end{equation} 
Notice the very strong 'level repulsion' (i.e., the absence of eigenvalues at small separations):  $P_n(s)=0$ for $0<s<(n+1)/2$.  For illustration, a few such examples are presented in Fig.~\ref{fig_old_map}(b). 

As in previous Section consider a particular case with 
\begin{equation}
\theta_j=\frac{2\pi j}{N}\, ,\qquad j=0,\ldots,N-1\, .
\end{equation}
Direct calculations shows that in this case matrix $M_{nm}$ denote by $z_{nm}$ takes the form
\begin{equation}
z_{nm}=\frac{1}{N\sin(\pi(m-n+1/2)/N)}\, .
\end{equation}
The matrix 
\begin{equation}
c_{nm}=e^{i\phi_n}z_{nm} ,\qquad n,m=1,\ldots,N
\label{c_matrix}
\end{equation}
is a  unitary matrix and in the limit $N\to\infty$ its spectral statistics tends to the shifted Poisson distribution \eqref{shifted_poisson}. 

It is these matrices that enter in the construction of block matrix  $C$  in \eqref{block_C} for even $N$. It is plain that in the limit $N_0\to\infty$ the $C$-matrix has the same asymptotic form as the above two models.  Due to the block structure of the $C$-matrix \eqref{block_C}, the square of its eigenvalues are eigenvalues of the product of two independent $c$ matrices
\begin{equation} 
\xi_{nm}=\sum_k e^{i\phi_n} z_{n k} e^{i\phi^{\prime}_k} z_{m k} 
\label{xi_matrix}
\end{equation} 
where $\phi_n$ and $\phi^{\prime}_k$ are two independent sets of $N_0$ random variables uniformly distributed between $0$ and $2\pi$. 

\begin{figure}
\begin{center}
\includegraphics[width=.7\linewidth]{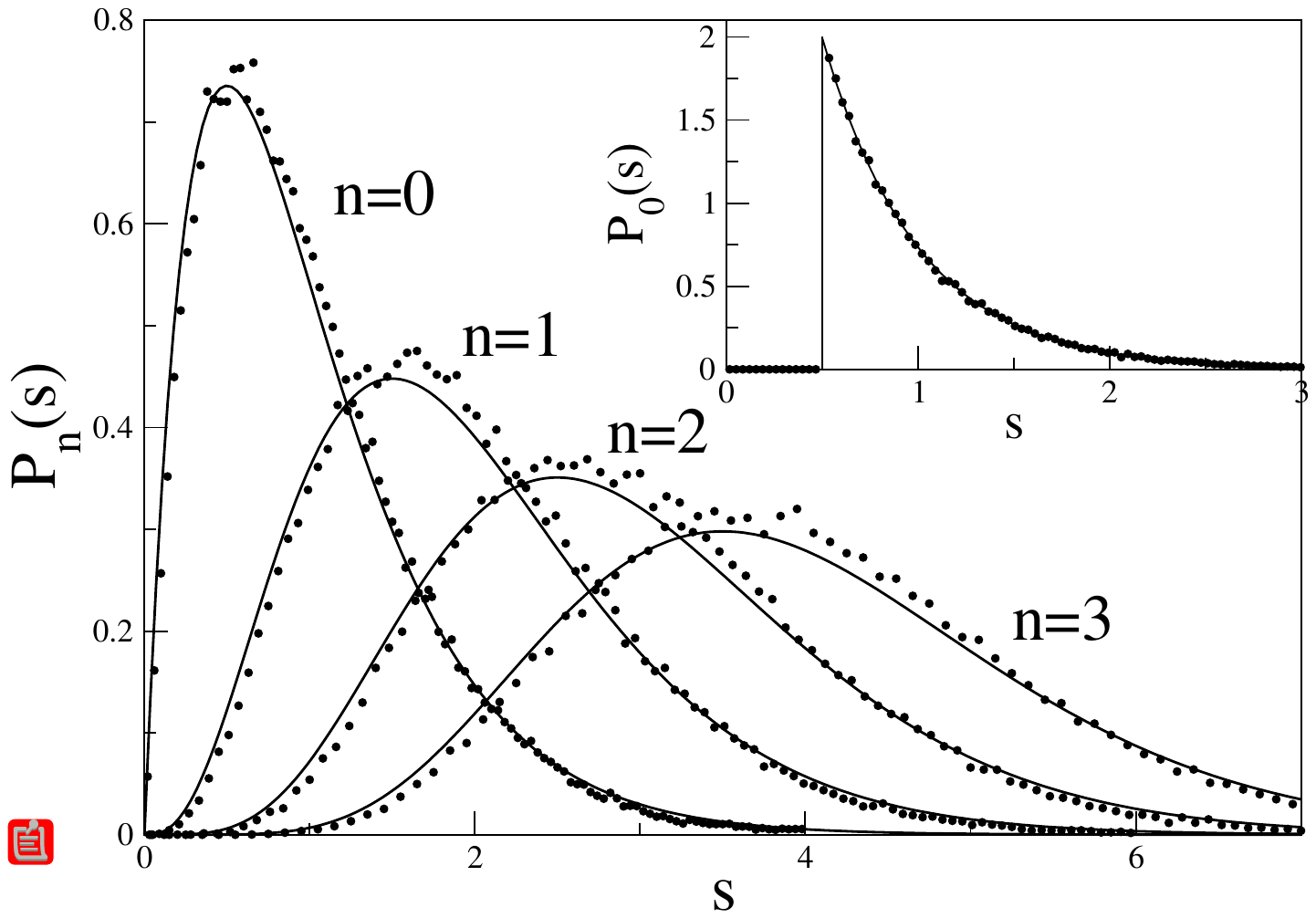}
\end{center}
\caption{The nearest-neighbour distributions $P_n(s)$ with $n=0,1,2,3$ for the product of two independent $c$-matrices \eqref{xi_matrix}. In the calculations the matrix dimension is $N_0=300$ and $300$ realisations of random phases were taken into account. Solid lines are the semi-Poisson expressions \eqref{semi_poisson}: $P_0(s)=4se^{-2s}$, $P_1(s)=8s^3e^{-2s}/3$, $P_2(s)=8s^5 e^{-2s}/15$, and $P_3(s)=16 s^7 e^{-2s}/315$. Insert: the nearest-neighbour distribution for one  $c$-matrix \eqref{c_matrix}. The calculations were done for $N_0=300$ and $300$ realisations of random phases were used. Solid line is the theoretical prediction \eqref{shifted_poisson}: $P_0(s)=0$ for $0<s<1/2$ and $P_0(s)=2e^{-2s+1}$ for $s>1/2$. }
\label{product_distribution}
\end{figure}

According to the above asymptotic equivalence conjecture spectral statistics of matrix $\xi$ has to be the semi-Poisson one.  The results of numerical calculations of the nearest-neighbour distributions for this matrix are presented in Fig.~\ref{product_distribution}. It is clearly seen that it is close to the semi-Poisson values \eqref{semi_poisson}. For completeness, in the Insert of this figure the nearest-neighbour distribution of one $c$-matrix is presented together with the theoretical prediction \eqref{shifted_poisson}.      

Careful  discussion of spectral properties of products of certain matrices with intermediate statistics will be given somewhere.    

\section{Trace formula for barrier billiards}\label{trace_formula}

Trace formulas are  the main tool in the semiclassical approach to quantum problems \cite{gutzwiller}. Such formulas relate the density of quantum eigenenergies to a sum over classical periodic orbits of  a given system. For pseudo-integrable billiards periodic orbits are organised   in families of parallel rays restricted from the both sides by billiard edges and the trace formula for such models has the form \cite{richens_berry}
\begin{equation}
d(E)=\bar{d}(E)+d^{(\mathrm{osc})}(E)
\end{equation}
where
\begin{equation}
\bar{d}(E)=\frac{\mathcal{A}}{4\pi},\qquad \mathcal{A}=\mathrm{billiard\;area}
\end{equation}
and 
\begin{equation}
d^{(\mathrm{osc})}(E)=\frac{1}{4\pi}\sum_{p}\frac{\epsilon_p \mathcal{A}_p}{\sqrt{2 \pi k \,L_p}}\exp\big ( ik\,L_p -i\pi/4\big ) +\mathrm{c.c.}\, .
\end{equation}
Here the summation is done over all classical periodic orbits. $L_p$ is the periodic orbit length,  
$\mathcal{A}_p$ is the area occupied by a given periodic orbit family, and $\epsilon_p$ is the phase factor related with billiard boundary conditions.  

Only for special pseudo-integrable billiards these quantities can be calculated analytically (cf., \cite{schmit}) but for the barrier billiard indicated in Fig.~\ref{semiclassical_BB}(b) their computation is straightforward. From Appendix~\ref{ppo}  (see also Appendix D in \cite{olivier_thesis}) it follows that for a primitive periodic orbit  
\begin{eqnarray}
L_{p}&=&\sqrt{(2aM)^2+(2bN)^2}\, , \qquad M,N=\mathrm{co-prime\;integers} \, ,\\
\epsilon_p \mathcal{A}_p&=&(-1)^{K} \big (1-2 \eta  \big )4ab \, ,\qquad K=\left [\frac{Nh_1}{a}\right ]\, ,\qquad \eta=\left \{\frac{Nh_1}{a}\right \} \, .
\label{ppo_contribution}
\end{eqnarray} 
Within the transfer operator formalism the density of state can be expressed by standard arguments. From the quantisation condition \eqref{quantisation} one  concludes that formally (with $k=k+i0$) 
\begin{equation}
d^{(\mathrm{osc})}(E)=-\frac{1}{2\pi i }\frac{\partial}{\partial E} \mathrm{Tr}\ln(\det(1+B(k+i0)))=\frac{1}{4\pi i k } \frac{\partial}{\partial k}\sum_{r=1}^{\infty} \frac{(-1)^r}{r} \mathrm{Tr}\,B^r(k+i0)+\mathrm{c.c.}\, .
\end{equation}
The difference between the transfer operator discussed in the paper and the general theory of quantum maps (see, e.g.,\cite{uzy})  lies in the definition of periodic orbits. In quantum maps periodic orbits are built  from  phases  in the transfer matrices  but in the barrier billiards  periodic orbits are fixed geometrically (cf. \eqref{ppo_contribution}) and are not directly related with phases of the $B$-matrix \eqref{B_matrix}.  As usual, the periodic orbit lengths for barrier billiards will reappear after the saddle-point summation but the determination  of a non-trivial pre-factor proportional to the area occupied by a periodic orbit, $\mathcal{A}_p$ in \eqref{ppo_contribution} within the transfer operator approach, seems,  has not been discussed in the literature.  The rest of this Section is devoted to such calculations. 
 
The periodic orbits contribution is contained in terms with even powers of the $B$-matrix $B^{2M}$ (as only in such terms one can get $2a M$ terms).  The trace includes  the summation over even and odd indices. In Section~\ref{odd_indices} it is demonstrated how to sum over all odd indices (which corresponds to  waves started from the Dirichlet part  of the billiard boundary) with fixed even indices. It is evident that the  full trace is twice this value.  

It will be shown below  that  for large $k$ 
\begin{equation}
\mathrm{Tr}\,B^{2M} (k+i0)=\sum_{n=1}^{\infty} F(n) e^{2ia M  p_{2n}} \, ,\qquad p_{2n}=\sqrt{k^2-\frac{\pi^2 n^2}{b^2}}
\label{partial_contribution}
\end{equation}
where function $F(n)$ is even: $F(-n)=F(n)$. 

Using  the Poisson summation formula (corrections at small $n$ can be added if necessary)
\begin{equation}
\sum_{n=1}^{\infty} F(n) \exp(-2i p_n a r)=
\frac{1}{2}\sum_{N=-\infty}^{\infty} \int_{-\infty}^{\infty}F(n) e^{-i\Phi(n)}d\,n 
\end{equation}
where 
\begin{equation}
\Phi(n)=2\pi  N n +2  a M \sqrt{k^2-\frac{\pi^2 n^2}{b^2}} \, .
\end{equation}
For large $k$ this integral can be calculated by the saddle-point method. 
The saddle-point values of $n$ are  determined from the condition $\Phi^{\prime}(n)=0$ which gives
\begin{equation}
n_{\mathrm{s.p.}}=\frac{2b^2 N }{\pi L_p} k\, , \qquad \Phi(n_{\mathrm{s.p.}})=kL_p\, , \qquad p_{2n_{\mathrm{s.p.}}}=\frac{2a M}{L_p} k
\label{s_p_values}
\end{equation}
with $L_p=\sqrt{(2aM)^2+(2bN)^2}$.

It is plain  that   $\Phi(n_{\mathrm{s.p.}})$ is the action along  the periodic orbit  $(2aM,2bN)$, $ L_p$ is the length of this orbit, and  $p_{n_{\mathrm{s.p.}}}=k \frac{2ar}{L_p}$  is the momentum along this orbit. 

Finally the contribution of \eqref{partial_contribution} to the trace formula (a term with $N=0$ is requires a slightly different treatment)
\begin{equation}
d^{(\mathrm{osc})}(E)=\frac{ab}{\pi} \sum_{N=1}^{\infty} \frac{F(n_{\mathrm{s.p.}}) }{\sqrt{2 \pi kL_p} }e^{ikL_p-i\pi/4}+\mathrm{c.c.}\, .
\label{s_p_trace}
\end{equation}   
The explicit calculation of function $F(n_{\mathrm{s.p.}})$ is performed in Section~\ref{odd_indices} where it is shown that it is related to the trace of a certain matrix. Section~\ref{eigenvalues_Q} is devoted to the calculation of eigenvalues of this matrix which, in the end, permit to prove that the transfer operator method leads to the correct answer for the trace formula. 

\subsection{Summation over odd indices} \label{odd_indices}

Let us calculate the trace of powers of the $B$-matrix  when only forward scattering are taken into account because as was indicated above the reflected waves are small at large momenta.  One has 
\begin{equation}
\frac{1}{2}\mathrm{Tr}\, B^{2M}=\sum_{m_j } W_{m_1, m_2} W_{m_2, m_3}\ldots W_{m_{M}, m_1}
\label{Br}
\end{equation} 
where 
\begin{equation}
W_{m,n}=\sum_q B_{2m,2q-1}B_{2q-1,2n}=e^{ih_2 p_{2 m} } \sum_{q}\left ( S_{2m,2q-1} e^{2ih_1 p_{2q-1}}S_{2q-1,2n }\right ) e^{ih_2 p_{2 n } }\, .
\end{equation}
The factor $1/2$ appears because  the fixation of odd indices and summation over even ones give exactly the same answer due to the cyclic invariance of the trace. 

In the paraxial approximation \eqref{S_paraxial} one needs to calculate the following sum 
\begin{equation}
W_{m,n}=e^{ih_2 (p_{2 m}+ p_{2 n } )}\frac{1}{\pi^2} \sum_{q}\frac{1}{m-(q-1/2)} e^{2ih_1 p_{2q-1}}\frac{1}{n-(q-1/2)}\, .
\end{equation}
The presence of paraxial  $S$-matrix elements forces the summand $q$ to be close to external indices $m$ and $n$  
\begin{equation}
q=m+r,\qquad n=m+p,\qquad n=q+p-r\, .
\end{equation}
It will be shown that the summations over $r$  and $p$ converge which implies that for large $m$  $r$ and $p$  are small,  $r\ll m$ and $p\ll m$. Then the following expansions are justified 
\begin{equation}
p_{2q-1}=p_{2(m+r)-1}\approx p_{2m}-\pi x(r-1/2),\qquad 
p_{2n}=p_{2(m+p)}\approx p_{2m}-\pi x p,\qquad x=\frac{\pi m}{b^2 p_{2m}}\, .
\end{equation}
After these approximations one can sum over all $r$ which leads to the following expression  (as $h_1+h_2=a$)
\begin{equation}
W_{m,n}=e^{i a p_{2 m}+ia p_{2n} }g(n-m)
\end{equation}
where ($p=n-m$) and $y=x h_1$
\begin{equation} 
g(p)= \frac{e^{i\pi y p} }{\pi^2} \sum_{r=-\infty}^{\infty} \frac{1}{(r-1/2)(r-1/2-p)}\exp \big (-i2 \pi y(r-1/2)) \big)\, .
\end{equation}
When $p=0$
\begin{equation}
g(0)=\frac{1}{\pi^2} \sum_{r_1=-\infty}^{\infty}\frac{e^{-i2\pi y(r-1/2)}}{(r-1/2)^2}\, . 
\end{equation}
For $p\neq 0$ 
\begin{equation}
g(p)=-\frac{e^{i\pi y p}}{\pi^2} \sum_{r=-\infty}^{\infty}\left ( \frac{1}{r-1/2}-\frac{1}{r-1/2-p}\right )\frac{e^{-i2\pi y(r-1/2)}}{p} \, .
\end{equation}
Shifting $r\to r+p$ in the second term gives  
\begin{equation}
g(p) =-\frac{2i\sin(\pi y p)}{\pi^2 p } \sum_{r=-\infty}^{\infty}\frac{e^{-i2\pi y(r-1/2)}}{r-1/2} \, .
\end{equation}
The summation in the above expressions corresponds to the sum over odd integers. As 
\begin{equation}
\sum_{\mathrm{odd}}=\sum_{\mathrm{all}}-\sum_{\mathrm{even}}
\end{equation}
the remaining sums can be calculated from known identities for the Bernoulli polynomials  (see, e.g.,  \cite{bateman})
 \begin{equation}
 \sideset{}{'}\sum_{r=-\infty}^{\infty} \frac{e^{2 i \pi x r}}{r^2} =
 2\pi^2 B_2(x) , \qquad B_2(x)=\{x\}^2-\{x\}+\frac{1}{6}
 \end{equation}
 and 
 \begin{equation}
\sideset{}{'}\sum_{r=-\infty}^{\infty}\frac{e^{2\pi i x r}}{r}=-2\pi i B_1(x),\qquad B_1(x)=\{x\}-\frac{1}{2} 
\end{equation}
Split $y$ into the sum of integer, $[y]$, and fractional, $\{ y\}$, parts:  $y=[y]+\{y\}$ with $0<\{y\}<1$ then
\begin{equation}
\sum_{r=-\infty}^{\infty}\frac{1}{(r-1/2)^2} e^{2i \pi  y (r-1/2)}= 2 \pi^2(-1)^{[y]}(4 B_2(\{y\}/2)  -B_2(\{y\}))=\pi^2(-1)^{[y]}(1-2\{y\}) 
\end{equation}
and 
\begin{equation}
\sum_{r=-\infty}^{\infty}\frac{e^{2\pi i x(r-1/2)}}{r-1/2} =2(-1)^{[y]}(-2\pi i) \Big ( B_1(\{x\}/2)-\frac{1}{2}B_1(\{x\})\Big ) =i\pi (-1)^{[y]} \, .
\end{equation}
Finally
\begin{equation}
W_{m,n}=e^{i a (p_{2 m}+ p_{2 n } )}(-1)^{[y]}f_{m-n}(y)
\label{asym_W}
\end{equation}
where $y=x h_1$ and 
\begin{equation}
f_{m-n}(y) =\left \{\begin{array}{cc} 1-2\{y\}, &m=n\\
-\dfrac{2\sin(\pi  y (m-n))}{\pi (m-n)}, &m\neq n \end{array}. 
 \right .
 \label{fmn}
\end{equation}
Put the obtained expression \eqref{asym_W} into  general formula \eqref{Br} 
and assume that all $m_j$ are close to each other
\begin{equation}
 p_{2m_j}\approx p_{2m_1}-\pi x  (m_j-m_1) ,\qquad x= \frac{\pi m_1}{b^2p_{2m_1}}\, .
\end{equation}
In such approximation there is only one large index $m_1$ and $M-1$ relatively small differences $m_{j}-m_1$. Notice that $x$ and, consequently, $y=x h_1$ depend only on $m_1$.  In the end one gets 
\begin{equation}
\frac{1}{2}\mathrm{Tr}\, B^{2M}=\sum_{m_1=1}^{\infty}  e^{2i a M p_{2m_1}} F(m_1)
\label{B_2M}
\end{equation}
where 
\begin{equation}
F(m_1)=(-1)^{M[y]} \sum_{m_2}\ldots \sum_{m_M}  f_{m_1-m_2}(y) f_{m_2-m_3}(y)\ldots f_{m_{M}- m_1}(y) e^{i\Phi_M}\, .
\label{F_m}
\end{equation}
The total phase $\Phi_M$ in this expression is 
\begin{equation}
\Phi_M=2a \sum_{j=1}^M  p_{2m_r}-2a M p_{2m_1 } \approx  -2 a\pi x \sum_{j=2}^{M} (m_j-m_1).
\label{Phi_M}
\end{equation} 
The sum in  \eqref{B_2M} is exactly as in \eqref{partial_contribution} and according to \eqref{s_p_trace}  its contribution to the trace formula  in the saddle-point approximation reduces to the calculation of the value of $F(m_1)$ at $m_1$ equals the saddle-point  value (see \eqref{s_p_values}) 
\begin{equation} 
m_{\mathrm{s.p.}}=\frac{2b^2 N }{\pi L_p} k, \qquad p_{2m_{\mathrm{s.p.}}}=\frac{2a M}{L_p} k
\end{equation}
It implies that in the saddle-point approximation 
\begin{equation}
x\equiv \frac{\pi m_{\mathrm{s.p.}} }{b^2 p_{2m_{\mathrm{s.p.}} }}  =\frac{N}{a M}\, .
\label{x_s_p}
\end{equation}
The phase factor $\Phi_M$ in  \eqref{Phi_M} takes the following value
 \begin{equation}
\Phi_M=- \frac{ 2\pi N}{M} \Big(\sum_{j=2}^M m_j-(M-1)m_1\Big)=
- \frac{ 2\pi N}{M} \sum_{j=1}^M m_j +2\pi N m_1\, .
\end{equation}
As $\Phi_M$ enters in the exponent (see \eqref{F_m})  the term $2\pi N m_1$  can be omitted. 

From this  one concludes that the function $F_{\mathrm{s.p.}}$ is 
\begin{equation}
F_{\mathrm{s.p.}}=(-1)^{M [y_{\mathrm{s.p.}}]}  (Q^M)_{m_{\mathrm{s.p.}},m_{\mathrm{s.p.}}}
\label{F_s_p}
\end{equation}
where 
\begin{equation}
Q_{mn}=e^{-2\pi i z m} f_{m-n}(y), \qquad z=\frac{N}{M},\qquad y=y_{\mathrm{s.p.}}=z\frac{h_1}{a} \, .
\label{Qmn}
\end{equation}
By the construction function $F_{\mathrm{s.p.}}$ is independent on $m_1$ therefore   the above equation can be rewritten in the form
\begin{equation}
F_{\mathrm{s.p.}}=(-1)^{M [y_{\mathrm{s.p.}} ]} \lim_{R\to\infty} \frac{1}{R}\mathrm{Tr}\, (Q^M)
\label{trace_Q}
\end{equation}
with $Q_{mn}$ in \eqref{Qmn} considered as $R\times R$ matrix. 

\subsection{Eigenvalues of the $Q$-matrix}\label{eigenvalues_Q}

To calculate the trace in \eqref{trace_Q} it is necessary to find eigenvalues of $R\times R$ matrix $Q_{m,n}$  in the limit $R\to\infty$. 

Consider a matrix  
\begin{equation}
U_{m n}(y)=\frac{\sin(\pi y (m-n))}{\pi (m-n)}\, ,\qquad U_{m m}(y)= y\, ,\qquad m,n=1,\ldots, R \, . 
\label{Umn}
\end{equation}
This matrix at finite $R$ determine the so-called discrete prolate spheroidal sequences whose properties were thoroughly investigated in \cite{slepian}. 

The structure of eigenvalues and eigenfunctions of this matrix in the limit of $R\to\infty$ can be found from the following simple considerations. Let us consider the identity (with an odd integer $w$)
\begin{equation}
\sum_{\alpha=-(w-1)/2}^{(w-1)/2}\frac{e^{2\pi i \alpha (m-n)/R}}{R}=
\frac{\sin(\pi y (m-n)}{R\sin (\pi (m-n) /R)},\qquad y=\frac{w}{R} \, . 
\label{identity}
\end{equation}
Functions (Greek letters are used to numerate  eigenvalues)
\begin{equation}
\phi_m(\alpha)=\frac{e^{2\pi i \alpha m/R}}{\sqrt{R}}
\label{eigenfunctions}
\end{equation}
are orthogonal when $m,\alpha \in [1,R]$   and the above expression signifies that the matrix in the right-hand side of Eq.~\eqref{identity} has $w=yR$ eigenvalues equal $1$ and $(1-y)R$ zero eigenvalues. For large $R$ the matrix in \eqref{identity} tends to that in \eqref{Umn}. Therefore, for $R\to\infty$ matrix $U_{mn}$ with $0<y<1$ also has asymptotically $yR$ eigenvalues $+1$ and  $(1-y)R$  zero eigenvalues (plus $\mathcal{O}(\ln R)$ transitional eigenvalues between $0$ and $1$ \cite{slepian}). 

This all is valid when  $0<y<1$. The matrix is periodic in $y$ with period $2$. When $1<y<2$ one has $y=1+\{y\}$. Therefore if one defines $U_{mm}=\{y\}$ then with this definition 
  $U_{mn}(y)=(-1)^{m-n}U_{mn}(\{y\})$. As this matrix is conjugate to $U_{mn}(\{y\})$ it has the same eigenvalues. Finally $R\times R$ matrix 
\begin{equation}
\hat{U}_{mn}(y)=\frac{\sin(\pi y (m-n))}{\pi (m-n)}\, ,\qquad \hat{U}_{m m}(y)= \{y\}
\end{equation} 
for all $y$ has asymptotically  $\{y\}R$ unit eigenvalues and $(1-\{y\})R$ zero eigenvalues. 
  
$R\times R$  matrix $f_{m-n}(y)$ in \eqref{fmn} can be rewritten as follows 
\begin{equation}
f_{m-n}(y)\equiv \delta_{mn}-2 \hat{U}_{mn}(y) 
\end{equation}
and, consequently,  matrix $f_{m-n}(y)$  has asymptotically $\{y\} R$ eigenvalues equal $-1$ and $(1-\{y\})R$ eigenvalues equal $+1$. 

Let $e(\alpha)$ with $\alpha=1,\ldots,R$ denote eigenvalues of this matrix.
One can arrange all $R$ eigenvalues in such a way that the first $\{y\}R$ eigenvalues are $-1$ and the rest are equal to $+1$. As we are working in the limit $R\to\infty$ it is convenient to use a (quasi) continuous variable $\zeta=\alpha/N$. Then the function $e(\zeta)$  can be written in the following form
\begin{equation}
e(\zeta)=\left \{ \begin{array}{cc} -1\, ,& 0 < \zeta <\{y\}\\+1\, ,&\{y\}<\zeta<1\end{array}
\right . .
\label{e_zeta}
 \end{equation}
Consider now the matrix \eqref{Qmn} in the saddle-point approximation
\begin{equation}
Q_{mn}=e^{-2\pi i z m} f_{m-n}(y)\, , \qquad z=\frac{N}{M}\, ,\qquad y=z\frac{h_1}{a}\, .
\label{Qz_mn}
\end{equation}
The phase in this expression depends only on the fractional part of $z=N/M$: $\{z\}=n/M$ with $n<M$.  For  primitive periodic orbits  $M$ and $N$ are co-prime integers, therefore, $n$ and $M$ are also co-prime integers. 

The key point in the investigation of eigenvalues of matrix $Q$ is the observation that any eigenfunction of matrix $f(m-n)$,  $\phi_m(\alpha)$,  in   \eqref{eigenfunctions} after the multiplication by phase factor $e^{-2\pi i z j m}$ with integer $j=\mathcal{O}(1)$ tends at large $R$ to another eigenfunction of the same matrix $f(m-n)$ with, in general, a different eigenvalue. Indeed  
\begin{equation}
e^{-2\pi i j n/M\,m} \phi_m(\alpha)=  \frac{e^{2\pi i m (\alpha- j n R/M)/R}}{\sqrt{R}}
\approx \phi_{m}(\beta)\, ,\qquad \beta=\alpha-v_j \mod R,\qquad   
v_j \equiv [j n R/M]\,  .
\end{equation}
One has  
\begin{equation}
\frac{j n}{M} R=\left [ \frac{j n}{M} R\right ]+\delta, \qquad \delta=\left \{\frac{j n}{M} R\right \}<1\, . 
\end{equation}
Therefore, the shift in the above expression differs from $v_j$ by  a small amount 
$\mathcal{O}(1)$ in  comparison to $\beta=\mathcal{O}(R)$ and (heuristically) can be ignored in the limit  $R\to\infty$. 

In the continuous approximation when variable $\zeta=\alpha/R \in (0,1)$ is introduced (as in \eqref{e_zeta}) such shift corresponds to  
\begin{equation}
\zeta \to  \{\zeta-\gamma_j\}\,  ,\qquad  \gamma_j=\left \{ j\, \frac{n}{M}\right \}\, .  
\end{equation}
Notice that for primitive periodic orbits integers $M$ and $N$ are co-prime and, consecutively.  the sequence $\{j\,n/M\}$ with $j=0,1,\ldots,M-1$  wraps all fractions $k/M$ with $k=0,1,\ldots, M-1$.  It means that the sequence of of $M$ eigenfunctions  $\phi_m(\zeta),\phi_m(\zeta-\gamma_1),\phi_m(\zeta-\gamma_2),\ldots,\phi_m(\zeta-(M-1)\gamma_{M-1})$ is closed under  multiplications by  $e^{-2\pi i n m/M} $.  

These considerations show that an eigenfunction of matrix $Q_{mn}$ in \eqref{Qmn} can in the limit $R\to\infty$  be approximated by by the sum 
\begin{equation}
\Phi_m =\sum_{j=0}^{M-1} c_j \phi_m(\zeta-\gamma_j) 
\end{equation}
where $\zeta$ is any number in the interval $(0,1/M)$. 
As in the continuous notations  
\begin{equation}
f_{m-n}(y)\phi_n(\zeta)=\epsilon_{\zeta} \phi_m(\zeta),\qquad \epsilon_{\zeta}=\pm 1 
\end{equation}
the eigenvalue condition 
\begin{equation}
Q_{mn}\Phi_n=\lambda \Phi_m
\end{equation}
implies that 
 \begin{equation}
 \sum_{j=0}^{M-1} c_j \epsilon_{\zeta-\gamma_j} \phi_m(\zeta -\gamma_{j+1} )=\lambda\left  (\sum_{j=0}^{M-1} c_j \phi_m(\zeta-\gamma_j)\right  )\, .
 \end{equation}
 Functions $\phi_m(\alpha)$ are orthogonal and this equation is equivalent to the following system of equations
 \begin{equation}
  \lambda c_0=\epsilon_{M-1}c_{M-1},\qquad \lambda c_1=\epsilon_0 c_0\, , \qquad \lambda c_2=\epsilon_1 c_1,\qquad  \ldots\, ,  \qquad
 \lambda c_{M-1}=\epsilon_{M-2} c_{M-2}\,  .
\end{equation}
 Therefore, new eigenvalues are determined from the condition (here $\epsilon_j\equiv \epsilon(\zeta-\gamma_j)$)
 \begin{equation}
 \lambda^M=\epsilon_0 \epsilon_1\ldots \epsilon_{M-1}=\prod_{j=0}^{M-1}\epsilon(\zeta-j/M)\, .
 \label{lambda_M}
 \end{equation}  
 As all $\epsilon_j$ are equal to either $+1$ or $-1$, $\lambda^M=\pm1$ which means that  new eigenvalues are strongly degenerated and can take only $2M$ different values (from the total number $R\to\infty$) 
 \begin{equation} 
 \lambda_{\beta}=e^{2\pi i \beta/(2M)}\, ,\qquad \beta=0,\ldots, 2M-1\, .
 \end{equation}
 As for the prolate spheroidal sequences the number of eigenvalues which differ from the above values is small with respect to the matrix dimension $R$ which agrees well with numerical calculations. 
 
To find the main quantity \eqref{F_s_p}  it is necessary to know  the value
\begin{equation}
\mathrm{Tr}\,Q^M=\sum \lambda^M 
\end{equation}
where the sum is taken over all eigenvalues of matrix $Q$. From \eqref{lambda_M} it follows that almost all eigenvalues in the $M^{\mathrm{th}}$ power in the limit $R\to\infty$ can take only two values $+1$ or $-1$  depending on the sign of the product of $\epsilon(\zeta- j/M)$ over $j=0,\ldots, M-1$. It is convenient to rename variables: $\zeta=\zeta-(M-1)/M$. 
To avoid the double counting new variable $\zeta$ has to be chosen between $0$ and $1/M$.  

From \eqref{e_zeta} one gets
\begin{equation}
\lim_{R\to\infty}\frac{1}{R}\mathrm{Tr}\,Q^M=M\int_{0}^{1/M}\left (\prod_{j=0}^{M-1} e(\zeta+j/M )\right ) d\zeta =\int_{0}^{1}\left (\prod_{j=0}^{M-1} e\Big (\frac{u+j}{M} \Big)\right ) d u\, .
\end{equation}
According to \eqref{e_zeta} 
\begin{equation}
e\Big (\frac{u+l}{M} \Big)=\left \{ \begin{array}{cc} -1, & u+l<M\{y\} \\+1 , & u+l>M\{y\}\end{array}\right . .
\end{equation}
 Split $M\{y\}$ with $y=\frac{Nh_1}{Ma}$ into integer and fractional parts
 \begin{equation}
 M\{y\}=s+\eta\, ,\qquad s=[M\{y\}]\, ,\qquad \eta=\{M\{y\}\}\, . 
 \end{equation}
Then it is plain that 
 \begin{equation}
 \prod_{l=0}^{M-1} e\Big (\frac{u+l}{M} \Big)=\left \{ \begin{array}{cc} s+1, & 0<u<\eta  \\ 
  s,  & \eta <u <1\end{array}\right .  
 \end{equation} 
 which using \eqref{trace_Q} proves that 
 \begin{equation}
 F_{\mathrm{s.p.}}=(-1)^{M [y]} (-1)^{s}(1-2\eta)\,  .
 \end{equation}
 As  $\{y\}=y-[y]$ one gets 
 \begin{equation}
 \eta\equiv \{M\{y\}\}=\{M y\}=\left \{ \frac{Nh_1}{a}\right \} \, ,\qquad s\equiv [M\{y\}]=M [y]-K,\qquad K=[ M y]\, .
 \end{equation}
 Finally 
 \begin{equation}
 F_{\mathrm{s.p.}}=(-1)^K(1-2\eta)\, ,\qquad K=\left [ \frac{N h_1}{a}\right ]\, ,\qquad \eta=\left \{ \frac{N h_1}{ a} \right \}
 \end{equation}
 which  agrees with the direct calculation of the contribution of a primitive periodic orbit  (cf. \eqref{ppo_contribution}) (the factor $4ab$ has been included in \eqref{s_p_trace}). 
  
 \section{Conclusion} 
 
 The semiclassical limit of high energy of the transfer operator for  symmetric barrier  billiards proposed in \cite{BB_I} is calculated and two related topics are investigated. 
 
 The first  concerns the spectral statistical properties of barrier billiards. By comparison of the asymptotic form of the transfer operator with the asymptotic of a certain class of random unitary matrices with known spectral statistics one concludes (at least heuristically) that spectral correlation functions of barrier billiards have to be described by the semi-Poisson formulas in a good agreement with numerical calculations.  
 
 The second subject consists in the derivation of the trace formula for barrier billiards within the transfer operator approach. Usually the semiclassical trace formulas are  obtained from standard geometrical considerations  but in the discussed method the calculations are not straightforward and require the knowledge of eigenvalues of certain unusual matrices.  Albeit  the answer was known in advance, the calculations reveal a new mechanism by which corrections to a saddle-point are combined to produce the pre-factor for the barrier billiard trace formula. 
 
Though the calculations start from  the exact transfer operator expression, it is demonstrated that  the semiclassical limit of many quantities could be obtained from simple physical considerations which may open a way of analytical investigations of more general pseudo-integrable models and,  in particular, of triangular billiards.        

\appendix

\section{Asymptotic calculations of  $K_{+}(\alpha)$}\label{asymptotic}
 
 From  Eq.~\eqref{K} it follows that $K_+(\alpha)$ can be rewritten as two convergent products
\begin{equation}
K_+(\alpha)=\Pi_1 \Pi_2\, , \qquad \Pi_1=\prod_{n=1}^{\infty}\Big (1-\frac{1}{2n}  \Big )e^{1/2n}\, ,\qquad \Pi_2= \prod_{n=1}^{\infty}\frac{p_{2n}+\alpha}{p_{2n-1}+\alpha}e^{-1/2n}\, .
\end{equation}
The first product is known \cite{bateman}
\begin{equation}
\Pi_1=\frac{e^{\gamma/2}}{\sqrt{\pi}}\, , \qquad 
\gamma=\lim_{N\to\infty}\left ( \sum_{n=1}^{N}\frac{1}{n}-\ln N\right ) \, .
\label{gamma}
\end{equation}
Here $\gamma$ is Euler's constant.  

The second product can be expressed as follows (where it is implicit  that $N\to\infty$)
\begin{equation}
\ln \Pi_2 =\sum_{n=1}^N\Big ( \ln \big (\sqrt{k^2-\pi^2 n^2/b^2}+\alpha\big ) -\ln\big (\sqrt{k^2-\pi^2 (n-1/2)^2/b^2}+\alpha\big ) -\frac{1}{2n}\Big )\, .
\end{equation}
This sum can be estimated  by  the Euler-Maclaurin  summation formula
\begin{equation}
\sum_{i=n+1}^m f(i)=\int_n^m f(x)dx+\frac{1}{2}\big (f(m)-f(n)\big )+
\sum_{p=1} \frac{B_{2p}}{(2p)!}\big (f^{(2p-1)}(m)-f^{(2p-1)}(n)\big )
\label{E_M}
\end{equation}
where $B_{2p}$ are the Bernoulli numbers $B_2=1/6$, $B_4=-1/30$, $B_6=1/42$.

Though the integrals are elementary,  for our purpose  it is sufficient  to approximate  the difference for large $k$  and/or large $n$  by the first term of expansion 
\begin{equation}
\ln \Pi_2 \approx \sum_{n=1}^N \Big (- \frac{n}{2\sqrt{k^2-\pi^2 n^2/b^2}\big  (\sqrt{k^2-\pi^2 n^2/b^2}+\alpha\big )}-\frac{1}{2 n} \Big ) .
\end{equation}
The summation over $n$ is  substituted by two integrals one from $0$ to $bk/\pi$ and the second from $bk/\pi$ to $N$ which corresponds, respectively,  to propagating and evanescent modes (in the last term  \eqref{gamma} is used)
 \begin{equation}
\ln \Pi_2 \approx -\frac{1}{2} (J_1+J_2)-\frac{1}{2}(\ln N+\gamma)
\end{equation}
where 
\begin{eqnarray}
J_1&=&\int_{0}^{b/\pi k}  \frac{(\pi^2/b^2)  n d n }{\sqrt{k^2-\pi^2 n^2/b^2}\big  (\sqrt{k^2-\pi^2 n^2/b^2}+\alpha\big )}= \ln (k+\alpha)-\ln \alpha\, ,\\
J_2&=&\int_{bk/\pi}^N  \frac{(\pi^2/b^2) n d n }{i\sqrt{\pi^2 n^2/b^2-k^2}\big  (i\sqrt{\pi^2 n^2/b^2-k^2 }+\alpha\big )}= -\ln (\pi N/b) +\ln ( \alpha )-\frac{\pi}{2}i\, .
\end{eqnarray}
The sign of the imaginary part ($i\alpha$) is fixed by the requirement that all singularities of $K_+(\alpha)$ are in the lower half-plane.  

Combining all factors one gets 
\begin{equation}
K_+^{(\mathrm{as})}(\alpha)\approx \frac{e^{i\pi/4}}{\sqrt{b(k+\alpha)}}. 
\label{K_asymptotic}
\end{equation}
This asymptotic is valid for large $k$ and $\alpha=\mathcal{O}(k)<k$. The error is assumed to  be  $\mathcal{O}(1/k)$.  

\begin{figure}
 \begin{minipage}{.49\linewidth}
\begin{center}
\includegraphics[width=.9\linewidth]{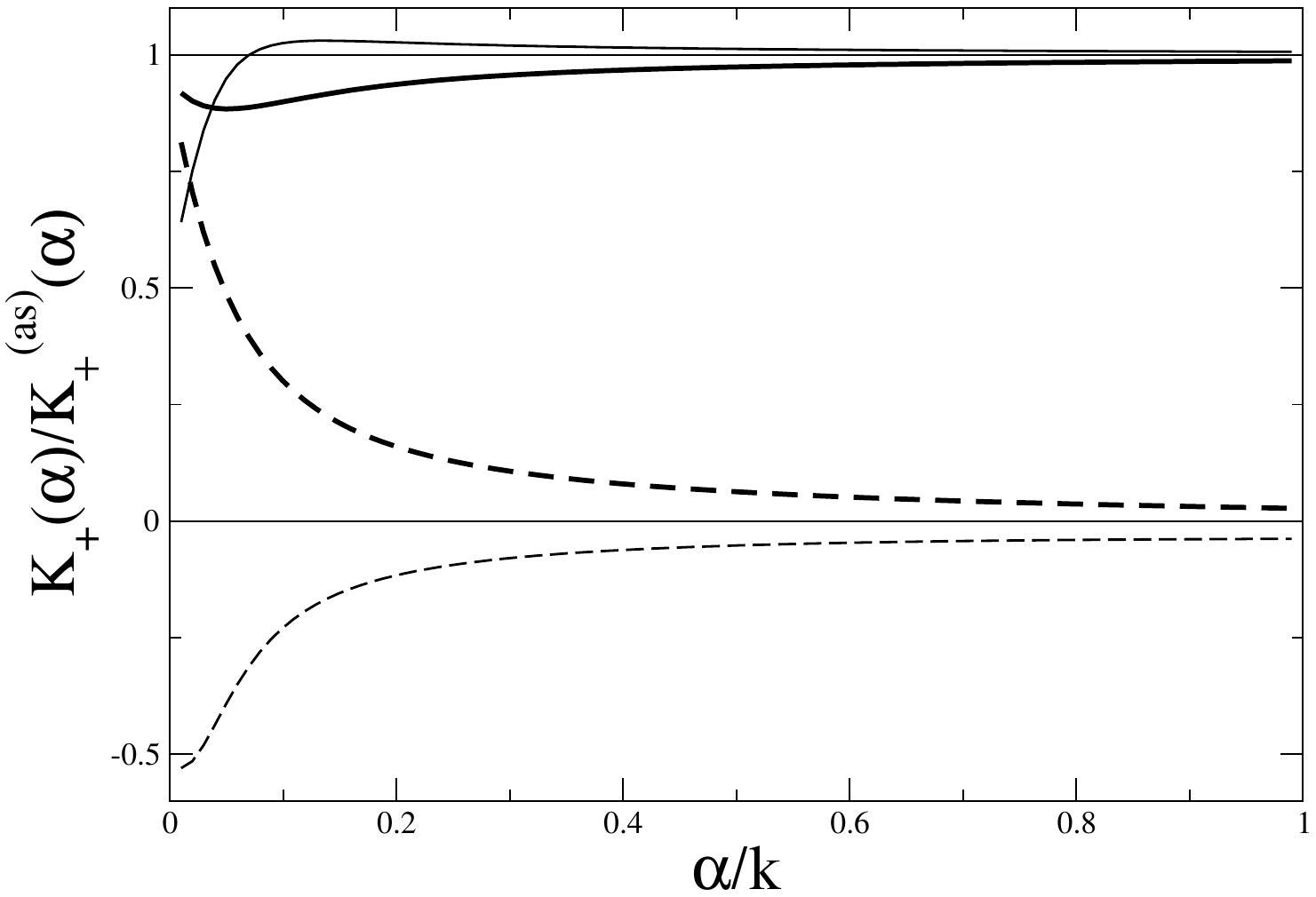}\\
a)
\end{center}
\end{minipage}
\begin{minipage}{.49\linewidth}
\begin{center}
\includegraphics[width=.9\linewidth]{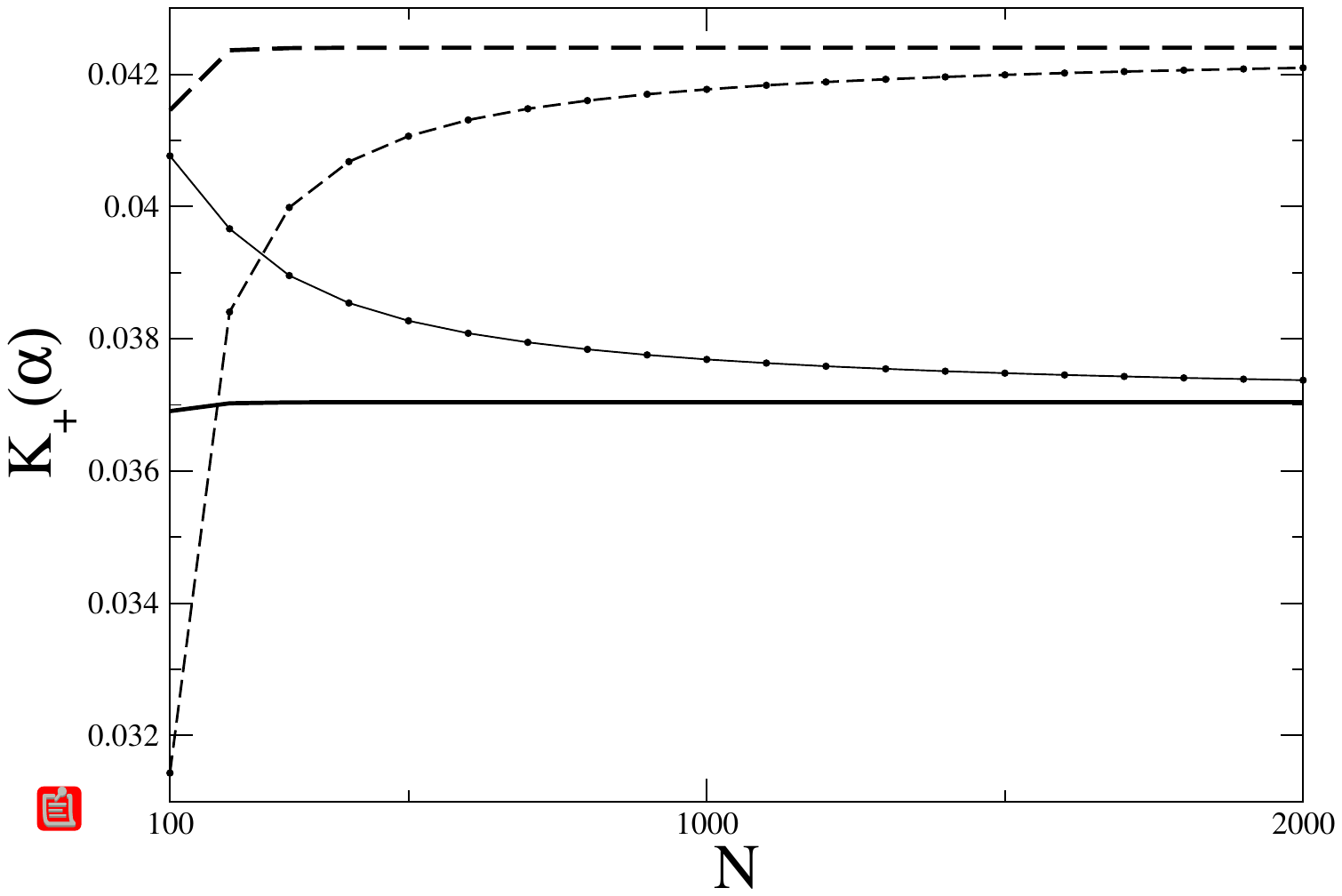}\\
b)
\end{center}
\end{minipage}
\caption{(a) The ratio of numerical value of $K_+(\alpha)$ calculated by  the multiplication of  $N=5000$ terms  with $b=1$ in \eqref{K} to  its asymptotic expression \eqref{K_asymptotic}. Solid lines - the real part of the ratio: thick line $k=200$ and thin line $k=500$. Dashed lines  - the imaginary part of the ratio with the same values of momenta. (b) $K_+(\alpha)$ versus the number of factors in \eqref{K}. Thin lines are result of the product of $N$ factors. Thick lines are  the results with correction terms \eqref{three_terms}. Solid (resp. dashed) lines indicate real and imaginary parts of  $K_+(\alpha)$. In the calculations $b=1$, $k=200$, and $\alpha=100.56$.}
\label{ratio}
\end{figure}

The ratio of $K_+(\alpha)$ calculated numerically to the asymptotic formula \eqref{K_asymptotic} is presented  in  Fig.~\ref{ratio}(a).  As expected, for not  small values of the argument this ratio indeed is close to $1$. Calculation of the behaviour of $K_+(\alpha)$ at small $\alpha/k$ is beyond the scope of the paper. 
 
As the transfer operator is exact it can be used also for numerical computations of the billiard barrier  spectrum. In such a case it is of interest to find a convenient method of calculation of $K_+(\alpha)$.  
By construction, the infinite product in \eqref{K}  converges and may be calculated simply by the multiplication of a large number, $N$, of factors (as it is done in Fig.~\ref{ratio}(a)). But the error  will be  only  $\mathcal{O}(1/N)$. Better accuracy can be achieved by  the approximation of  the rest by  the Euler-Maclaurin formula \eqref{E_M}. One has
\begin{equation}
K_{+}(\alpha)=\prod_{n=1}^{N }\Big ( 1-\frac{1}{2n} \Big ) \Big (\frac{p_{2n}+\alpha}{p_{2n-1}+\alpha}\Big )\, \exp\big (S_{\mathrm{cor}}(N)\big )\, .
\label{K_plus_correction}
\end{equation}
The correction  term, $S_{\mathrm{cor}}(N)$,  has the form (it is assumed that $N>bk/\pi$ to fix the correct branches)
\begin{equation}
S_{\mathrm{cor}}(N)=\sum_{n=N+1}^{\infty} \Big (f(n)-f(n-1/2)\Big )\, ,\quad f(x)=\ln\big (\sqrt{x^2-\kappa^2}-v\big )-\ln (x)
\label{difference}
\end{equation}
where  $\kappa=\frac{bk}{\pi},\; v=i \frac{b\alpha}{\pi}$
.
Take a few terms of the expansion of function $f(n)$ at large $n$
\begin{equation}
f(n)=\frac{a_2}{n^2}+\frac{a_3}{n^3}+\frac{a_4}{n^4}+\mathcal{O}(n^{-5})
\end{equation}
where
\begin{equation}
a_2=\frac{v}{2}\, ,\qquad a_3=\frac{2\kappa^2+2v^2+v}{4}\, ,\qquad a_4=\frac{6\kappa^2 v+4v^3+3\kappa^2+3v^2+v}{8}
\end{equation}
and calculate the necessary sums by the Euler-Maclaurin formula  \eqref{E_M}. In this way one obtains 
\begin{equation}
S_{\mathrm{cor}}(N)=\frac{a_2}{N}+\frac{a_3-a_2}{2N^2} +\frac{2a_4-3a_3+a_2}{6 N^3}+
\mathcal{O}(N^{-4})\, . 
\label{three_terms}
\end{equation}
At Fig.~\ref{ratio}(b)  it is shown  that adding just the above three correction terms amplifies drastically the accuracy. 

\section{Asymptotic of the $S$-matrix by method of images}\label{images}

Asymptotic formula \eqref{K_asymptotic} permits to find the behaviour of the full propagating $S$-matrix \eqref{S_matrix} at large momenta
\begin{eqnarray}
& & S_{2n-1,2m}=S_{2m,2n-1}= \frac{(-1)^{m+n} \pi m \sqrt{k+p_{2n-1}}}{b^2 (p_{2n-1}-p_{2m})\sqrt{p_{2n-1} p_{2m} (k+p_{2m})}}\, ,\label{s_2n-1_2m} \\
& & S_{2n-1,2m-1}=-i \frac{(-1)^{m+n} \sqrt{ (k+p_{2n-1})(k+p_{2m-1})}}{b (p_{2n-1}+p_{2m-1}) \sqrt{p_{2n-1} p_{2m-1}}}\ ,\label{s_2n-1_2m-1}\\
& & S_{2n,2m}=-i\frac{(-1)^{m+n} \pi^2 nm}{b^3 (p_{2n}+p_{2m}) \sqrt{p_{2n} p_{2m} (k+p_{2n})(k+p_{2m})}}\, . \label{s_2n_2m}
\end{eqnarray}
The purpose of this Appendix is to demonstrate that these asymptotic values can be obtained from the  well-known Sommerfeld solution for the diffraction on a half-plane \cite{sommerfeld, optics} .
 
 Formally, for any diffraction problem  the  wave at large distances from the diffraction centre is the  sum of an incident wave and a reflected wave. In two dimensions 
 \begin{equation}
\Psi(r,\theta)=e^{ikr\cos(\theta-\theta_i)}+\frac{1}{\sqrt{8\pi kr}}D(\theta_i,\theta ) e^{ikr -3\pi i /4}
\end{equation}
where $D(\theta_i,\theta)$ is a diffraction coefficient depended on the incident angle $\theta_i$ and the angle of reflection $\theta$. 

\begin{figure}
\begin{center}
\includegraphics[width=.4\linewidth]{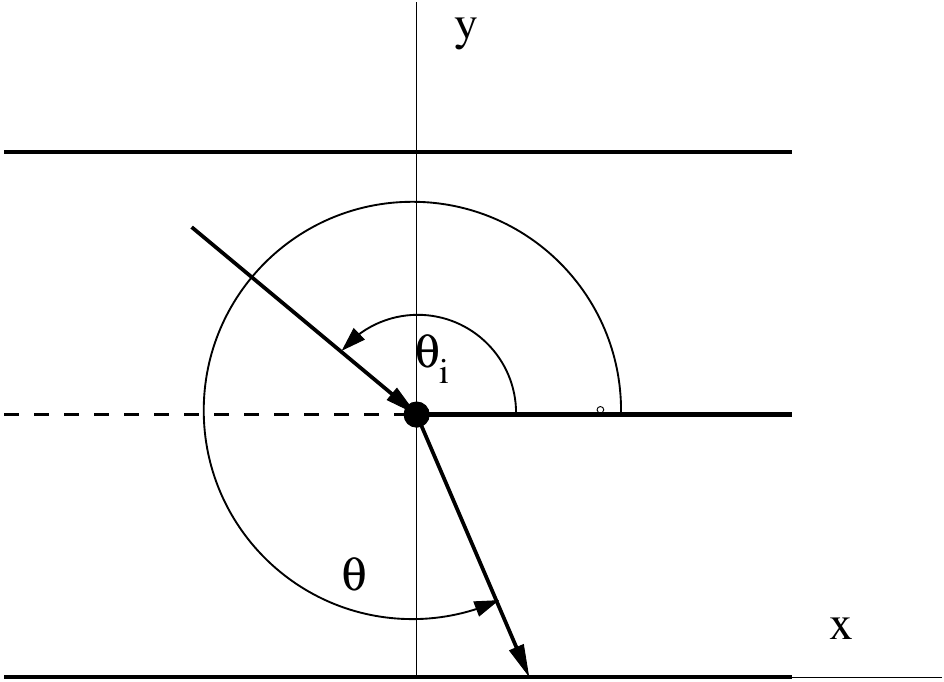}
\end{center}
\caption{Angles determining the Sommerfeld diffraction coefficient. } 
\label{sommerfeld_angles}
\end{figure}

For the scattering on a barrier (i.e., a half-plane) Sommerfeld shows \cite{sommerfeld,optics} that  
\begin{equation}
D(\theta_i, \theta)=\frac{1}{\cos\big ((\theta-\theta_i)/2\big )}-\frac{1}{\cos\big ((\theta+\theta_i)/2\big )}
\label{D_sommerfeld}
\end{equation}
where all angles  are calculated  counterclockwise from the barrier (see Fig~\ref{sommerfeld_angles})
. 

A principal difficulty in treating pseudo-integrable billiards consists in the impossibility to split scattered waves into the sum of initial and final waves which manifests in  the divergence of  the corresponding diffraction coefficient  in certain directions (called optical boundaries). For diffraction coefficient as in \eqref{D_sommerfeld} optical boundaries appear at $\theta=\pi\pm \theta_i$.  Waves in vicinities of optical boundaries are complicated which lead to many  unusual properties of pseudo-integrable billiards (cf. \cite{superscars}).

The normalised incident wave for the barrier problem  in the cartesian coordinates $(x,y)$ as in Fig.~\ref{sommerfeld_angles} is  indicated in \eqref{elementary_solutions}.
This wave can be represented as the sum of two plane waves with  incident angles
\begin{equation}
\theta_i=\pi \pm \phi\, ,\qquad \cos\phi=\frac{p_{2n-1}}{k}\, .
\label{theta_i}
\end{equation}
In a vicinity of the barrier tip the scattered wave takes the form
\begin{equation}
\Psi^{(\mathrm{sc})}(r,\theta)=\frac{C}{\sqrt{8\pi kr}}D(\phi,\theta ) e^{ikr -3\pi i /4} 
\end{equation}
where $C$ is the value of the incident wave at the barrier tip (i.e.,  $x=0,y=b$ in Fig.~\ref{sommerfeld_angles})
\begin{equation}
C= -\frac{(-1)^n}{\sqrt{b p_{2n-1}}} 
\end{equation} 
and $D(\phi,\theta)$ is the diffraction coefficient \eqref{D_sommerfeld} with substitution \eqref{theta_i} 
\begin{equation}
D(\phi,\theta )=\frac{1}{\sin((\theta+\phi)/2)}+\frac{1}{\sin((\theta-\phi)/2)}\, ,\qquad \cos(\phi)=\frac{p_{2n-1}}{k}\, . 
\end{equation}
After the scattering on the barrier tip there are two possibilities. If $\theta>3\pi /2$ the wave will propagate between two horizontal boundaries with the Dirichlet boundary conditions.  If 
$\pi<\theta<3\pi/2$ the wave has to propagate in a tube with the Dirichlet  condition on the lower  boundary and the Neumann boundary on the upper boundary.  In the both cases the field in a point $(x,y)$ with $0<y<b$ is the sum of all rays which connect  the barrier tip with this point after multiple reflections on horizontal boundaries. These reflections correspond  to all unfolded images of the final point:  $y_r=\pm y +2b r $ with integer $r$. 

Therefore, the distances between the tip and the final points  are 
\begin{equation}
R_r^{(\pm)}=\sqrt{x^2+(b\pm y+2b r ))^2}\, ,\qquad   \cos\big (\theta_r^{(\pm )}\big )=\frac{x}{R_r^{(\pm)}}\, .
\end{equation}
Here $\theta_r$ is the smallest angle between the straight line passed through  the barrier   and the corresponding classical ray.  

The total wave is the sum over all these rays.  The transmitted field with $x>0$ is 
\begin{equation}
\Psi^{(\mathrm{tr})}(x,y)=\sum_{r} \Big ( \Psi^{(\mathrm{sc})} (R_r^{(-)}, \theta_r^{(-)})-\Psi^{(\mathrm{sc})} (R_r^{(+)}, \theta_r^{(+)})\Big )\, .
\end{equation}
It is clear that $\Psi^{(\mathrm{tr})}(x,0)=\Psi^{(\mathrm{tr})}(x,b)=0$. The reflected field with $x<0$  is given by a sum   
\begin{equation}
\Psi^{(\mathrm{ref})}(x,y)=\sum_r (-1)^r \Big ( \Psi^{(\mathrm{sc})}
 (R_r^{(-)}, \theta_r^{(-)})- \Psi^{(\mathrm{sc})} (R_r^{(+)}, \theta_r^{(+)})\Big )
 \label{psi_r}
\end{equation}
which obeys different boundary conditions: $\Psi^{(\mathrm{ref})}(x,0)=0$ and $\partial_y \Psi^{(\mathrm{ref})}(x,b)=0$. 

The usual way of calculating such sums is to use the Poisson summation formula and the saddle-point method. For completeness, these calculations are present below.

Let $R(r)=\sqrt{x^2+(z+2b r )^2}$ then 
\begin{equation}
\sum_{r=-\infty}^{\infty} \frac{1}{\sqrt{8\pi k R(r)}} e^{ik R(r)}=
\sum_{m=-\infty}^{\infty} \int_{-\infty}^{\infty} \frac{1}{\sqrt{8\pi k R(r)}} e^{i\Phi(r)}dr\, , \qquad \Phi(r)=R(m)-2\pi r m\, .
\end{equation} 
The saddle-point value of $r$ is determined from the condition $\Phi^{\prime}(r^{(\mathrm{sp})})=0$ which leads to 
\begin{eqnarray}
&&r^{(\mathrm{sp})}=-\frac{z}{2b}+\frac{\pi m x}{2b^2 \sqrt{k^2-\pi^2 m^2/b^2}}\, , \qquad 
R(r^{(\mathrm{sp})})=\frac{k x}{\sqrt{k^2-\pi^2 m^2/b^2}}\, , \qquad  \Phi(r^{(\mathrm{sp})})=\frac{\pi z}{b}m\, ,\nonumber \\
 &&\Phi^{\prime\prime}(r^{(\mathrm{sp})})=\frac{4b^2 k x^2}{R^3(r^{(\mathrm{sp})})}\, ,\qquad 
\cos \theta(r^{(\mathrm{sp})})=2\pi - \tilde{\theta},\qquad \tilde{\theta}=\frac{\sqrt{k^2-\pi^2 m^2/b^2}}{k}\, . 
\end{eqnarray}
Finally in the saddle-point approximation
\begin{equation}
\sum_{r=-\infty}^{\infty} \frac{1}{\sqrt{8\pi k R(r)}} e^{ik R(r)}=\sum_{m} 
\frac{1}{4b \sqrt{k^2-\pi^2 m^2/b^2}} e^{ \left (i \pi z m/b +ik\sqrt{k^2-\pi^2 m^2/b^2}+i\pi/4  \right )}\, .
\label{green_function}
\end{equation}
This result implies that 
\begin{equation}
\Psi^{(\mathrm{tr})}(x,y)=\frac{(-1)^n}{2\sqrt{bp_{2n-1}}}\sum_{m=1}^{\infty}\frac{ (-1)^m e^{i x p_{2m}}}{b p_{2m}} D(\phi,\tilde{\theta}) \sin(\pi m y/b) \, .
\label{psi_tr}
\end{equation}
By definition, the $S$-matrix element $S_{2n-1,2m}$ is the  coefficient in front of the normalised final function $e^{i x p_{2m}}\sin(\pi m y/b)/\sqrt{b p_{2m}}$ (cf. \eqref{elementary_solutions})
\begin{equation}
S_{2n-1,2m}= \frac{(-1)^{m+n} D(\phi,\tilde{\theta}) }{2 \sqrt{b p_{2n-1}} \sqrt{b p_{2m}}}=
\frac{(-1)^{m+n} \sin(\tilde{\theta}) \cos(\phi/2)}{\sqrt{b p_{2n-1}} \sqrt{b p_{2m}} (\cos(\phi)-\cos(\tilde{\theta}))\cos(\tilde{\theta}/2)}\, .
\end{equation}
As $\cos \phi=p_{2n-1}/k$ and $\cos(\tilde{\theta})=p_{2m}/k$ this result coincides with \eqref{s_2n-1_2m}.  

The reflected wave can be obtained similarly. The differences with the above formulas are the following. First, due to the factor $(-1)^r$ in \eqref{psi_r} one has to substitute $m\to m-1/2$ which, in particular,  implies that Eq.~\eqref{psi_tr} has to be multiply by $-i$. Second, due to a different definition  $\theta=\pi + \theta^{\prime}$ with $\cos \theta^{\prime}=p_{2m-1}/k$ and 
 \begin{equation}
 D(\phi,\pi+\tilde{\theta})=\frac{4 \cos(\tilde{\theta}/2)\cos(\phi/2)}{\cos \tilde{\theta}-\cos\phi }\, .
 \end{equation}
Finally
\begin{equation}
S_{2n-1,2m-1}=-i \frac{2(-1)^{m+n} \cos(\tilde{\theta}/2)\cos(\phi/2) }
{\sqrt{b p_{2n-1}} \sqrt{b p_{2m-1}} (\cos(\phi)+\cos(\tilde{\theta}))}
\end{equation}
which is the same as in \eqref{s_2n-1_2m-1}.

\section{Contribution of a primitive periodic orbit in barrier billiards}\label{ppo}

Periodic orbits in barrier billiards form parallel periodic families (pencils) restricted from the both sides by singular points.  Let us consider one periodic orbit  which connects the origin (i.e., low-left corner of barrier billiard) with point of coordinates $(2Ma, 2Nb)$ (see Fig.~\ref{semiclassical_BB}(b)) where $M$ and $N$ are two integers. For primitive orbits these integers  have to be co-prime.  In Fig.~\ref{semiclassical_BB}(b) it means that 
\begin{equation}
\tan \phi=\frac{Nb}{Ma}
\end{equation}
and the equation of this orbit is 
\begin{equation}
y=\frac{Nb}{Ma} x\, .
\end{equation}
Singular vertices in the considered billiard (see Fig.~\ref{semiclassical_BB}(b)) after unfolding are situated in points with coordinates 
\begin{equation}
(\pm h_1+2ma, (2n+1)b)
\end{equation} 
 with integers $m$ and $n$. 
 
 It is well known (and can easily be checked by direct calculations) that the distance between a point with coordinates $(x_0,y_0)$ and a line determined by the equation $Ax+By+D=0$ is 
 \begin{equation}
\Delta=\frac{A x_0+B y_0+D}{\sqrt{A^2+B^2}}\, . 
 \end{equation} 
(Different signs of $\Delta$ correspond to points in different sides of the line.)

Therefore the distances between all singular points and the periodic orbit passed through the origin are
\begin{equation}
\Delta_{\pm}=\frac{Ma (2n+1)b-Nb(\pm h_1+2 m a)}{\sqrt{(Ma)^2+(Nb)^2}} \, .
\end{equation}
Denote  $L_p=\sqrt{(2Ma)^2+(2Nb)^2}$,  $\omega_p =2ab/L_p $, and separating the integer and fractional parts of $Nh_1/a$ 
\begin{equation}
N \frac{h_1}{a} =K+\eta\, ,\qquad K=\left [N \frac{h_1}{a} \right ]\, ,\qquad \eta=\left \{N  \frac{h_1}{a}\right \}
\end{equation}
one gets 
\begin{equation}
\Delta_{\pm}=\omega \big ( 2(Mn -Nm) +M \mp (K+\eta)  \big )\, . 
\end{equation}
As integers $M$, $N$ are co-prime  there exist two integers $\mu$ and $\nu$ such that $M\nu-N\mu=1$. Therefore if $M$ and $K$ are of the same parity  the minimum of the distances are equal to
\begin{equation}
\Delta_{\pm}=\mp \omega_p \eta
\end{equation}
and if $M$ and $K$ are of different parity then the minimal  distances are 
\begin{equation}
\Delta_{\pm}=\pm (1-\eta)\, . 
\end{equation}
Finally the width of one periodic orbit channel (which includes an orbit through the origin) is 
\begin{equation}
\Delta_1=\left \{ \begin{array}{cc}  2\omega_p \eta\, , & M, K=\mathrm{the\;same\;parity}\\
2\omega (1-\eta)\, , & M, K=\mathrm{opposite\;parity}\end{array}\right . .
\end{equation} 
The total width of any periodic orbit family is $2\omega_p$ and the second periodic orbit pencil has the width $\Delta_2=2\omega-\Delta_1$. As two channels correspond to periodic orbits with different total signs one concludes that the contribution of the above primitive periodic orbit is  (taking into account the   factor $L_p$ due to the implicit integration over the longitudinal coordinate.)
\begin{equation}
g=-(-1)^{M+K} \phi_0 \big (1-2 \eta  \big )4ab 
\label{area_pp}
\end{equation} 
where $\phi_0$ is the phase factor (the Maslov index) of a periodic trajectory passed through the origin. 
This factor equals $(-1)^{n_c}$ where $n_c$ is the number of crossings the Neumann parts of the boundary by the considered orbit. For the orbit which starts from the origin it can be found without calculations by the following arguments. Such orbit passes through the point $(Ma,Nb)$ which is the centre of the rectangle with sides $2Ma$ and $2Nb$. It is plain that the bottom-left rectangle and the top-right rectangle are mirror images of each other.  Therefore, the total number of intersections with Neumann boundaries inside these rectangles will be even. The only possibility to have odd numbers is the crossing the Neumann boundary exactly in the centre with coordinates $(Ma,Nb)$ but it is  possible only when $M$ is even and $N$ is odd. As integers $M$ and $N$ are co-prime it means that when $M$ is  even $N$ has to be automatically odd. Therefore, the phase $\phi_0$ is the follows 
\begin{equation}
\phi_0=\left \{ \begin{array}{cc} 1, & M=\mathrm{odd}\\ -1 ,&M=\mathrm{even}\end{array}\right . =-(-1)^M \, .
\end{equation} 
Final formula is obtained from \eqref{area_pp}
\begin{equation}
g=(-1)^{K} \big (1-2 \eta  \big ) 4ab \, ,\qquad K=\left [\frac{Nh_1}{a}\right ]\, ,\qquad \eta=\left \{\frac{Nh_1}{a}\right \} \, . 
\label{final_area}
\end{equation} 
It is independent on $M$ and depends only on $Nh_1/a$.


\end{document}